\newcommand{\HI}{H\,{\sc {i}}}
\newcommand{\HII}{H\,{\sc {ii}}}
\newcommand{\MHI}{M$_{\rm H \sc I}$}
\newcommand{\NHI}{$N$(H\,{\sc {i}})}
\newcommand{\SFEHI}{SFE(H\,{\sc {i}})}
\newcommand{\TAUHI}{$\tau_{\rm H{\sc {I}}}$}
\newcommand{\SIGMAHI}{$\Sigma_{\rm H{\sc {I}}}$}

\newcommand{\halpha}{\ensuremath{{\rm H}\alpha}}
\newcommand{\lsim}{~\rlap{$<$}{\lower 1.0ex\hbox{$\sim$}}}
\newcommand{\gsim}{~\rlap{$>$}{\lower 1.0ex\hbox{$\sim$}}}

\documentclass[twocolumn]{aastex63}

\usepackage{multirow}
\usepackage{gensymb}

\graphicspath{{figure/}}

\accepted{July 30, 2020}
\submitjournal{ApJ}

\shorttitle{Dwarf-dwarf galaxy merger VCC 848}
\shortauthors{Zhang et al.}
\begin{document}

\title{The Blue Compact Dwarf Galaxy VCC 848 Formed by Dwarf-Dwarf Merging: \HI~Gas, Star Formation and Numerical Simulations}

\author[0000-0003-1632-2541]{Hong-Xin Zhang}
\affil{Key Laboratory for Research in Galaxies and Cosmology, Department of Astronomy, University of Science and Technology of China, Hefei, Anhui 230026, China}
\affil{School of Astronomy and Space Science, University of Science and Technology of China, Hefei, Anhui 230026, China}
\email{hzhang18@ustc.edu.cn}

\author[0000-0001-5303-6830]{Rory Smith}
%\affil{Korea Astronomy and Space Science Institute, Daejeon 305-348, Republic of Korea}
\affil{Korea Astronomy and Space Science Institute (KASI), 776 Daedeokdae-ro, Yuseong-gu, Daejeon 34055, Republic of Korea}
%\email{rorysmith@kasi.re.kr}

\author{Se-Heon Oh}
\affil{Department of Physics and Astronomy, Sejong University, 209 Neungdong-ro, Gwangjin-gu, Seoul, Republic of Korea}

\author[0000-0003-2922-6866]{Sanjaya Paudel}
\affil{Department of Astronomy and Center for Galaxy Evolution Research, Yonsei University, Seoul 03722, Republic of Korea}

\author{Pierre-Alain Duc}
\affil{Université de Strasbourg, CNRS, Observatoire astronomique de Strasbourg, UMR 7550, F-67000 Strasbourg, France}

\author{Alessandro Boselli}
\affil{Aix Marseille Université, CNRS, LAM (Laboratoire d’Astrophysique de Marseille) UMR 7326, F-13388 Marseille, France}

\author[0000-0003-1184-8114]{Patrick C\^ot\'e}
\affil{National Research Council of Canada, Herzberg Astronomy and Astrophysics Program, 5071 West Saanich Road, Victoria, BC V9E 2E7, Canada}

\author{Laura Ferrarese}
\affil{National Research Council of Canada, Herzberg Astronomy and Astrophysics Program, 5071 West Saanich Road, Victoria, BC V9E 2E7, Canada}

\author{Yu Gao}
\affil{Department of Astronomy, Xiamen University, Xiamen, Fujian 361005, China}
\affil{Purple Mountain Observatory, Chinese Academy of Sciences, 10 Yuanhua Road, Nanjing 210023, China}

\author{Deidre A. Hunter}
\affil{Lowell Observatory, 1400 West Mars Hill Road, Flagstaff, AZ 86001, USA}

\author{Thomas H. Puzia}
\affil{Instituto de Astrofísica, Pontificia Universidad Católica de Chile, 7820436 Macul, Santiago, Chile}

\author[0000-0002-2073-2781]{Eric W. Peng}
\affil{Department of Astronomy, Peking University, Beijing 100871, China}
\affil{Kavli Institute for Astronomy and Astrophysics, Peking University, Beijing 100871, China}

\author{Yu Rong}
\affil{Instituto de Astrofísica, Pontificia Universidad Católica de Chile, 7820436 Macul, Santiago, Chile}

\author{Jihye Shin}
\affil{Korea Astronomy and Space Science Institute (KASI), 776 Daedeokdae-ro, Yuseong-gu, Daejeon 34055, Republic of Korea}

\author{Yinghe Zhao}
\affil{Yunnan Observatories, Chinese Academy of Sciences, Kunming 650011, China}

%\author{Kaixiang Wang}
%\affil{Department of Astronomy, Peking University, Beijing 100871, China}
%\affil{Kavli Institute for Astronomy and Astrophysics, Peking University, Beijing 100871, China}

%\author{Butler Burton}
%\affiliation{National Radio Astronomy Observatory}
%\affiliation{AAS Journals Associate Editor-in-Chief}
%\nocollaboration



\begin{abstract}

A clear link between a dwarf-dwarf merger event and enhanced star formation (SF) {\em in the recent past} was recently identified in the gas-dominated merger remnant VCC 848, offering by far the clearest view of a gas-rich late-stage dwarf-dwarf merger.\ We present a joint analysis of JVLA \HI~emission-line mapping, optical imaging and numerical simulations of VCC 848, in order to examine the impact of the merger on the stellar and gaseous distributions.\ VCC 848 has less than 30\% of its \HI~gas concentrated within the central high-surface-brightness star-forming region, while the remaining \HI~is entrained in outlying tidal features.\ Particularly, a well-defined tidal arm reaches $N$(\HI) comparable to the galaxy center but lacks SF.\ The molecular gas mass inferred from the current SF rate (SFR) dominates over the atomic gas mass in the central $\sim$ 1.5 kpc.\ VCC 848 is consistent with being a main-sequence star-forming galaxy for its {\em current} stellar mass and SFR.\ The \HII~region luminosity distribution largely agrees with that of normal dwarf irregulars with similar luminosities, except that the brightest \HII~region is extraordinarily luminous.\ Our $N$-body/hydrodynamical simulations imply that VCC 848 is a merger between a gas-dominated primary progenitor and a gas-bearing star-dominated secondary.\ The progenitors had their first passage on a near-radial non-coplanar orbit more than 1 Gyr ago.\ The merger did not build up a core as compact as typical compact dwarfs with centralized starburst, which may be partly ascribed to the star-dominated nature of the secondary, and in a general sense, a negative stellar feedback following intense starbursts triggered at early stages of the merger.

\end{abstract}



\section{Introduction} \label{sec:intro}

In the standard $\Lambda$ cold dark matter ($\Lambda$CDM) paradigm, dark matter halos, together with visible galaxies sitting close to their centers, 
assemble hierarchically under the influence of gravity.\ 
Galaxy mergers, especially those involving gas-rich galaxies of comparable masses (i.e.\ ``wet'' major mergers with primary-to-secondary mass 
ratios $\lesssim$ 4), can dramatically change galaxy morphologies \citep[e.g.][]{toomre72, barnes91, mihos94}, enhance star formation activities 
\citep[e.g.][]{sanders96, mihos96, duc97, zhang10, luo14, cibinel19} and trigger active galactic nuclei \citep[e.g.][]{ellison11, treister12, weston17} 
in short timescales.\ Observational studies in the past decade have established that the galaxy merger rates increase steadily with redshift at least 
up to $z$ $\sim$ 2-3 \citep[e.g.][]{lotz11, lopezsanjuan15, mundy17, duncan19, ventou19}.\

Although galaxy merging events in the local universe are not as common as in the early universe, nearby galaxy mergers provide unique laboratories 
for detailed investigations of the influence of merging process on galaxy evolution.\ A vast majority of studies of galaxy mergers so far have focused 
on relatively massive galaxies, while mergers between dwarf galaxies (M$_{\star}$ $<$ 10$^{9}$ $M_{\odot}$) received little attention until very recently, which is 
partly due to the limited depth of most existing surveys.\ Being close to the bottom of the hierarchical structure formation, dwarf galaxy mergers are 
expected to be more common than their massive counterparts.\ Recent cosmological zoom-in simulations suggest that about 10\%-20\% of Local Group 
satellite galaxies with M$_{\star}$ $>$ 10$^{6}$ M$_\odot$ have experienced a major merger event since $z = 1$ \citep{deason14}.\

The first systematic study of gas-rich dwarf-dwarf interacting pairs was presented by \cite{stierwalt15}, who found that most of their isolated dwarf pairs are 
as gas-rich as unpaired dwarfs of similar stellar masses, and that the current star formation rate (SFR) of dwarf pairs is enhanced by a factor of $\sim$ 2 on 
average at relatively small projected separations, similar to what has been found for massive galaxy pairs \citep[e.g.][]{ellison13, silva18, pearson19}.\ 
\cite{pearson16} further showed that dwarfs involved in interacting pairs tend to have more extended atomic gas distribution than their unpaired analogues.\ 
By searching for low surface brightness merger signatures, \cite{paudel18a} compiled a catalog of 177 relatively low mass merger candidates 
(M$_{\star}$ $<$ 10$^{10}$ M$_\odot$), most of which turn out to be star-forming galaxies.\ \cite{kado20} found that 15\%-20\% of nearby dwarf galaxies with 
extreme starburst activities show signs of tidal debris.\ In addition to the above relatively systematic studies, a few more case studies of gas-rich dwarf pairs have 
also been carried out \citep{annibali16, privon17, paudel17b, paudel18b, makarova18, johnston19}.\

Previous studies of gas-rich dwarf pairs are biased toward systems at relatively early merging stages \citep[e.g.][]{stierwalt15}.\ It remains unclear  
how dwarf-dwarf merging events impact the overall star formation activities and re-shape the gaseous and stellar distributions of the 
merger remnants.\ Particularly, it has long been conjectured that many blue compact dwarf galaxies (BCD), which are characterized by having bluer colors, 
more intense star formation activities and unusually higher central surface brightness compared to ordinary dwarf irregular galaxies \citep[e.g.][]{gildepaz03}, 
might be formed through gas-rich dwarf-dwarf merging \citep[e.g.,][]{vanzee98, noeske01, ostlin01, bekki08, lelli12, lelli14, watts16}.\ Other plausible formation 
mechanisms for the compact stellar distributions of BCDs include inspiraling of giant star formation clumps driven by dynamical friction \citep{elmegreen12} and 
central starburst sustained by pristine gas accretion from the local environs or even the cosmic web \citep[e.g.][]{johnson12, lopezsanchez12, ashley14, verbeke14}.\ 
BCD galaxies are the closest local analogues to UV-luminous galaxies such as Lyman break galaxies and Ly$\alpha$ emitters detected at high redshift (e.g., 
\citealt{gawiser07, finkelstein11, shibuya19}).\ It is generally difficult to differentiate the above mentioned formation mechanisms for BCDs and other starburst dwarf 
galaxies, partly due to the often morphologically irregular appearance of star-forming dwarf galaxies.

In a recent work (\citealt{zhang20}, hereafter Paper I), we report a discovery of remarkably extended stellar shells around the BCD galaxy VCC 848 located in the outskirts of the Virgo 
cluster, based on deep optical imaging data from the Next Generation Virgo Cluster Survey (NGVS; \citealt{ferrarese12}).\ This discovery confirms that VCC 848 is the 
remnant of a merger between dwarf galaxies.\ VCC 848 is perhaps the clearest known example of a star-forming dwarf galaxy resulting from a gas-rich 
merger.\ Paper I shows that VCC 848 is likely formed by merging between two dwarf galaxies with comparable masses within a factor of a few, and the merging event 
has significantly enhanced the star formation activities in the past $\sim$ 1 Gyr.\ In the rest of the current paper, we present interferometric observations of \HI~gas in VCC 848, 
study the recent star formation activities and discuss the merging process by invoking numerical simulations.\ Some relevant properties of VCC 848, which are either from the 
literature or derived in this work, are summarized in Table \ref{propertysummary_tab}.\ Throughout this paper, we adopt a distance of 16.5 Mpc for VCC 848 \citep{mei07,blakeslee09} 
and use the \cite{schlafly11} Galactic extinction map to correct the photometry.\

%\Distribution of the best-fit surface brightness profile models.
%\begin{longrotatetable}
\begin{deluxetable*}{l c c}
\tabletypesize{\small}
\tablecolumns{3}
\setlength{\columnsep}{5pt}
%\tablewidth{68in}
\tablecaption{\label{propertysummary_tab} Properties of VCC 848}
\tablehead{
\colhead{Property}
&\colhead{Value}
&\colhead{Reference} \\
%\colhead{(1)}
%&\colhead{(2)}
%&\colhead{(3)} \\
\noalign{\vskip -4.5mm}
}

\startdata
%\noalign{\vskip -0.5mm}
%\hline
Other name\dotfill & A1223+06 & \nodata \\
Morphological classification   &  iI,M BCD  &  1 \\
Distance\dotfill & 16.5 Mpc & 2 \\
Heliocentric radial velocity\dotfill & 1532 km s$^{-1}$ & This work \\
Angular distance from M87\dotfill & 6.7$^{\degree}$ & \nodata \\
Angular distance from M49\dotfill & 2.4$^{\degree}$ & \nodata \\
Absolute $B$ magnitude\dotfill & $-16.05$ mag\tablenotemark{~b} & 1  \\
Absolute $g$ magnitude\dotfill & $-16.46$ mag & This work \\
($g-i$) color\dotfill & 0.46 mag & This work \\
12+log(O/H)\dotfill & 8.03 & 3 \\
%$i$-band half-light radius\dotfill & 2.80 kpc & This work \\
$i$-band scale length of stellar main body\dotfill & 0.67 kpc & This work \\
$\mu_{0,g}$\dotfill & 20.8 mag arcsec$^{-2}$\tablenotemark{~a} & This work \\
$\mu_{0,g}-\mu_{0,i}$\dotfill & 0.3 mag\tablenotemark{~a} & This work \\
Stellar mass\dotfill & $2.1\times10^{8}$ $M_{\odot}$ & This work \\
\HI~mass\dotfill & $4.2\times10^{8}$ $M_{\odot}$\tablenotemark{~b} & 4 \\
$V_{\rm max}$\dotfill & 41.5 km s$^{-1}$ & This work \\
Radius of $V_{\rm max}$\dotfill & 1.6 kpc & This work \\
\halpha~luminosity\dotfill & $3.9\times10^{39}$ ergs s$^{-1}$\tablenotemark{~b} & 1 \\
Total IR luminosity\dotfill & $4.0\times10^{7}$ $L_{\odot}$\tablenotemark{~b,c}  & 5 \\
SFR$_{{\rm FUV+IR}}$\dotfill & 0.025 $M_{\odot}$ yr$^{-1}$\tablenotemark{~d} & This work \\
SFR$_{\halpha+{\rm IR}}$\dotfill & 0.023 $M_{\odot}$ yr$^{-1}$\tablenotemark{~d} & This work \\
SFR$_{\rm IR}$\dotfill & 0.002 $M_{\odot}$ yr$^{-1}$\tablenotemark{~d} & This work \\
\hline
\enddata
\tablenotetext{a}{Central surface brightness and color measured within the central 1 arcsec in semi-major axis}
\tablenotetext{b}{Value has been adjusted to our adopted distance}
\tablenotetext{c}{Total IR luminosity derived by combining the {\it Herschel} 100 $\mu$m, 160 $\mu$m and 250 $\mu$m photometry of \cite{grossi15}, following the recipe of \cite{galametz13}}
\tablenotetext{d}{SFR derived from the observed FUV, H$\alpha$ or IR luminosity by adopting equations 8 and 10 of \cite{catalan15}}
%\tablecomments{
References: (1) \citealt{gildepaz03}; (2) \citealt{mei07,blakeslee09}; (3) \citealt{vilchez03,lee03,zhao13}; (4) \citealt{haynes11}; (5) \citealt{grossi15}
%}

\end{deluxetable*}

%\end{longrotatetable}

\section{Observations and Data Reduction} \label{sec:obs}

\subsection{Optical and Far-infrared Images}
Broadband $u$-, $g$-, $i$-, and $z$-band images of VCC 848 were obtained by NGVS with the MegaCam instrument on the Canada-France-Hawaii Telescope.\ 
With a dedicated data acquisition strategy and processing pipeline, the NGVS reaches a 2$\sigma$ surface brightness limit of $\mu_{g}\simeq29$ mag arcsec$^{-2}$.\ 
The processed NGVS images have a pixel scale of 0\farcs186.\ The full width at half maximums (FWHM) of the point spread function (PSF) are 0\farcs78, 0\farcs71, 
0\farcs53, and 0\farcs62, respectively, for the $u$, $g$, $i$, and $z$ passbands of the VCC 848 field.\ An in-depth analysis of these broadband images of VCC 848 
has been presented in Paper I.

Narrow-band \halpha~imaging of VCC 848 was obtained with the Las Campanas 100-inch du Pont telescope by \cite{gildepaz03}, as part of a large imaging 
observational campaign for nearby BCDs.\ The total exposure time for VCC 848 was 45 minutes and the seeing FWHM was $\sim$ 1\farcs4 (pixel scale $=$ 
0\farcs26).\ \cite{gildepaz03} did continuum subtraction of their narrow-band images by using the $R$ broadband images.\ We correct the continuum-subtracted 
fluxes for the contamination of [N {\sc ii}] lines by adopting a [N {\sc ii}]$\lambda$6584/H$\alpha$ line ratio of 0.069 determined by \cite{vilchez03}.\

Archival far-UV (FUV) image from the {\it Galaxy Evolution Explorer} ({\it GALEX}) All-sky Imaging survey is available for VCC 848.\ With an exposure time of 112s, 
the FUV emission is detected only in the central high surface brightness region of VCC 848, and the integrated FUV magnitude is 16.9 $\pm$ 0.05 \citep{voyer14}.\ 
We do not use the shallow FUV image in the work, except that we calculate an integrated SFR by combining the FUV (corrected for the Galactic extinction) and 
infrared (see below) photometry (Table \ref{propertysummary_tab}).\

{\it Herschel} Photodetector Array Camera and Spectrometer (PACS) 100 $\mu$m, 160 $\mu$m and Spectral and Photometric Imaging Receiver (SPIRE) 250 $\mu$m, 
350 $\mu$m, 500 $\mu$m observations of VCC 848 were presented by \cite{grossi15}, as part of the {\it Herschel} Virgo Cluster Survey (\citealt{auld13}).\ Among these 
far-infrared (FIR) images, the PACS 100 $\mu$m has the highest spatial resolution (FWHM = 9\farcs4), and more importantly has been shown to be the most reliable 
monochromatic estimator of total IR (TIR) luminosity among all of the $Spitzer$ or $Herschel$ wavebands, with a scatter of 0.05 dex \citep{galametz13}.\ We will therefore 
use the 100 $\mu$m image to evaluate the contribution of obscured star formation to the total star formation budget at local scales of VCC 848 .\

\subsection{JVLA \HI~Emission Line Mapping}

\subsubsection{Observations}
%\Distribution of the best-fit surface brightness profile models.
%\begin{longrotatetable}
\begin{deluxetable}{lcccc}
\tabletypesize{\footnotesize}
\tablecolumns{5}
\setlength{\columnsep}{-0.1pt}
\tablewidth{0pt}
\tablecaption{The JVLA \HI~Observations of VCC 848}
\tablehead{
\colhead{Array}
&\colhead{Date}
&\colhead{TOS\tablenotemark{~a}}
&\colhead{No.\ of Channels}
&\colhead{Ch Width} \\
\colhead{}
&\colhead{(yy-mm-dd)}
&\colhead{(hr)}
&\colhead{}
&\colhead{(km s$^{-1}$)}\\
\colhead{(1)}
&\colhead{(2)}
&\colhead{(3)}
&\colhead{(4)}
&\colhead{(5)} \\
\noalign{\vskip -4.5mm}
}

\startdata
\noalign{\vskip -0.5mm}
%\hline
B   &  2016-05-30  &  0.27  &  512  &  1.7\\
B   &  2016-07-02  &  0.17  &  512  &  1.7\\
B   &  2016-07-12  &  0.30  &  512  &  1.7\\
B   &  2016-07-21  &  0.30  &  512  &  1.7\\
B   &  2016-07-22  &  0.17  &  512  &  1.7\\
B   &  2016-07-29  &  1.20  &  512  &  1.7\\
B   &  2016-07-31  &  1.60  &  512  &  1.7\\
C   &  2016-03-06  &  2.04  &  512  &  1.7\\
D   &  2017-04-02  &  0.90  &  512  &  1.7\\
\enddata
\tablenotetext{a}{Time on source (TOS)}
%\tablecomments{
%Col 1: JVLA configuration; Col 2: observing date; Col 3: time on source; Col 4: number of channels; Col 5: channel width in km s$^{-1}$).
%}
\label{obslog_tab}

\end{deluxetable}
%\end{longrotatetable}

VCC 848 was observed in B, C, and D array configurations of the Jansky Very Large Array (JVLA; \citealt{perley11}) radio interferometer.\ 
The observations were performed in the 1-2 GHz L-band (primary beam $\simeq$ 30\arcmin) through multiple scheduling blocks (SBs) 
between 2016 May and 2017 Apr (Project ID: 16A-074; PI: Zhang; Table \ref{obslog_tab}).\ The JVLA WIDAR correlator was configured 
with two 1-GHz baseband IF pairs (A0/C0 and B0/D0) and the 8-bit sampler.\ The correlator integration time is 3 seconds.\ Each baseband 
was divided into 22 sub-bands with dual polarization products (RR and LL) each.\ Only one 4 MHz wide sub-band of the A0/C0 baseband 
was used for \HI~emission line observations presented in this paper.\ The 4 MHz sub-band was tuned to a central frequency corresponding 
to the systemic velocity of VCC 848.\ By using a recirculation factor of 4, the 4-MHz wide sub-band consists of 512 spectral channels, with a 
channel width of 7.81kHz ($\sim$1.7 km s$^{-1}$ for a rest frequency of 1420.4 MHz).

Each SB of our observations began and ended with a 10- to 15-minute scan on the primary flux calibrator 3C286.\ The primary flux 
calibrator was used to calibrate the flux scale and bandpass.\ In between the observations of the primary flux calibrator, the phase 
calibrator J1254+1141 was observed for 5-6 minutes before and after each scan (10-25 minutes) on our target galaxy.\ VCC 848 
was observed for a total on-source time of 4.0 hrs in the B array, 2.0 hrs in the C array, and 0.9 hrs in the D array (Table \ref{obslog_tab}).\

\subsubsection{Calibration}
The multi-configuration data sets from the spectral line sub-band were calibrated and imaged using the CASA software package \citep{mcmullin07}.\ 
The basic calibration procedure is as follows.\ (1) the raw science data from each observed SB were converted into a CASA Measurement Set (MS) 
using the task {\sc importevla} with online flags and other deterministic flags applied; (2) a phase reference antenna close to the central part of the array 
was chosen from either the east or west arms; (3) problematic data points were located ({\sc plotms}) and flagged ({\sc flagdata}); (4) derive the ITRF 
antenna position corrections ({\sc gencal}); (5) set the model visibility amplitude and phase of the flux calibrator ({\sc setjy}), and derives the initial phase 
and delay calibration of flux calibrator ({\sc gaincal}); (6) derive the bandpass calibration for the flux calibrator ({\sc bandpass}); (7) derive the complex 
gain calibrations for the flux and phase calibrators ({\sc gaincal}), and set the flux scale of phase calibrator ({\sc fluxscale}); (8) apply the above-generated 
various calibrations to MS of calibrators and target; (9) repeat steps 3-8 for the calibrated MS until no further flagging is needed.\

\subsubsection{Imaging}
Before combining the calibrated multi-configuration MSs using the CASA task {\sc concat}, the (relative) weight of visibilities 
in each calibrated MSs was first determined based on line-free channels ({\sc statwt}) and then the task {\sc cvel2} was used 
to transform channel coordinates to a common heliocentric velocity reference frame with a 1.7 km s$^{-1}$ channel width.\ 
Spectral continuum was subtracted from the combined MS ({\sc uvcontsub}) with a first-order polynomial fit to visibilities in 
the line-free channels.\

The continuum subtracted MS was deconvolved using {\sc tclean}, with a cell size of 1$\farcs$5 $\times$ 1$\farcs$5, a robust 
parameter of 0.7, and a circular clean region of 4\arcmin~in radius around the target.\ A robust parameter of 0.7 leads to a 
restoring beam size of 7\farcs1 $\times$ 6\farcs6 (PA = 56.78$^{\circ}$) and rms noise of 0.64 mJy beam$^{-1}$.\ The multi-scale 
clean algorithm was used for recovering emission structures on different spatial scales.\ After several tests, a set of scale sizes 
of 6\arcsec, 12\arcsec, 24\arcsec and 36\arcsec and a ``smallscalebias'' parameter of 0.7 were adopted for cleaning, down to a 
global stopping threshold of 2 times the rms noise.\ The residual maps were checked to make sure all the flux has been retrieved.

The \HI~data cube was blanked in order to separate features of genuine emission from pure noise, following similar procedures 
adopted in previous studies (e.g.\ \citealt{walter08}; \citealt{hunter12}).\ In particular, the original data cube was first smoothed 
to a spatial resolution of 15\arcsec$\times$15\arcsec, then pixels that belong to spatial features appearing in at least 3 consecutive 
channels and above 2$\times$ the rms noise of the smoothed cube were masked as area of real emission ({\sc imager.drawmask}).\ 
The blanking mask thus generated was applied to the original data cube ({\sc immath}).\ The blanked data cube at the original resolution 
will be used for the following analysis in this paper.

%\Distribution of the best-fit surface brightness profile models.
%\begin{longrotatetable}
\begin{deluxetable}{l c}
\tabletypesize{\footnotesize}
\tablecolumns{2}
\setlength{\columnsep}{5pt}
\tablewidth{5in}
\tablecaption{\label{himapsummary_tab} \HI~mapping parameters of VCC 848}
\tablehead{
\colhead{Parameters}
&\colhead{Value} \\
%\colhead{(1)}
%&\colhead{(2)}
%&\colhead{(3)} \\
\noalign{\vskip -4.5mm}
}

\startdata
%\noalign{\vskip -0.5mm}
%\hline
Briggs robustness parameter\dotfill   &  0.7 \\
Multiscale parameter\dotfill &   6$\arcsec$, 12$\arcsec$, 24$\arcsec$, 36$\arcsec$ \\
Pixel size\dotfill & 1\farcs5 \\
Synthesized beam\dotfill & 7\farcs1 $\times$ 6\farcs6 \\
Beam position angle\dotfill & 56.8$\degree$ \\
rms noise (mJy beam$^{-1}$ ch$^{-1}$)\dotfill & 0.64  \\
\NHI~rms noise ($\times$10$^{19}$ cm$^{-2}$ beam$^{-1}$ ch$^{-1}$)\dotfill & 1.5  \\
\hline
\enddata
%}

\end{deluxetable}

%\end{longrotatetable}

\subsubsection{Comparison with ALFALFA Observation}
\begin{figure}[]
\centering
\includegraphics[height=0.4\textwidth]{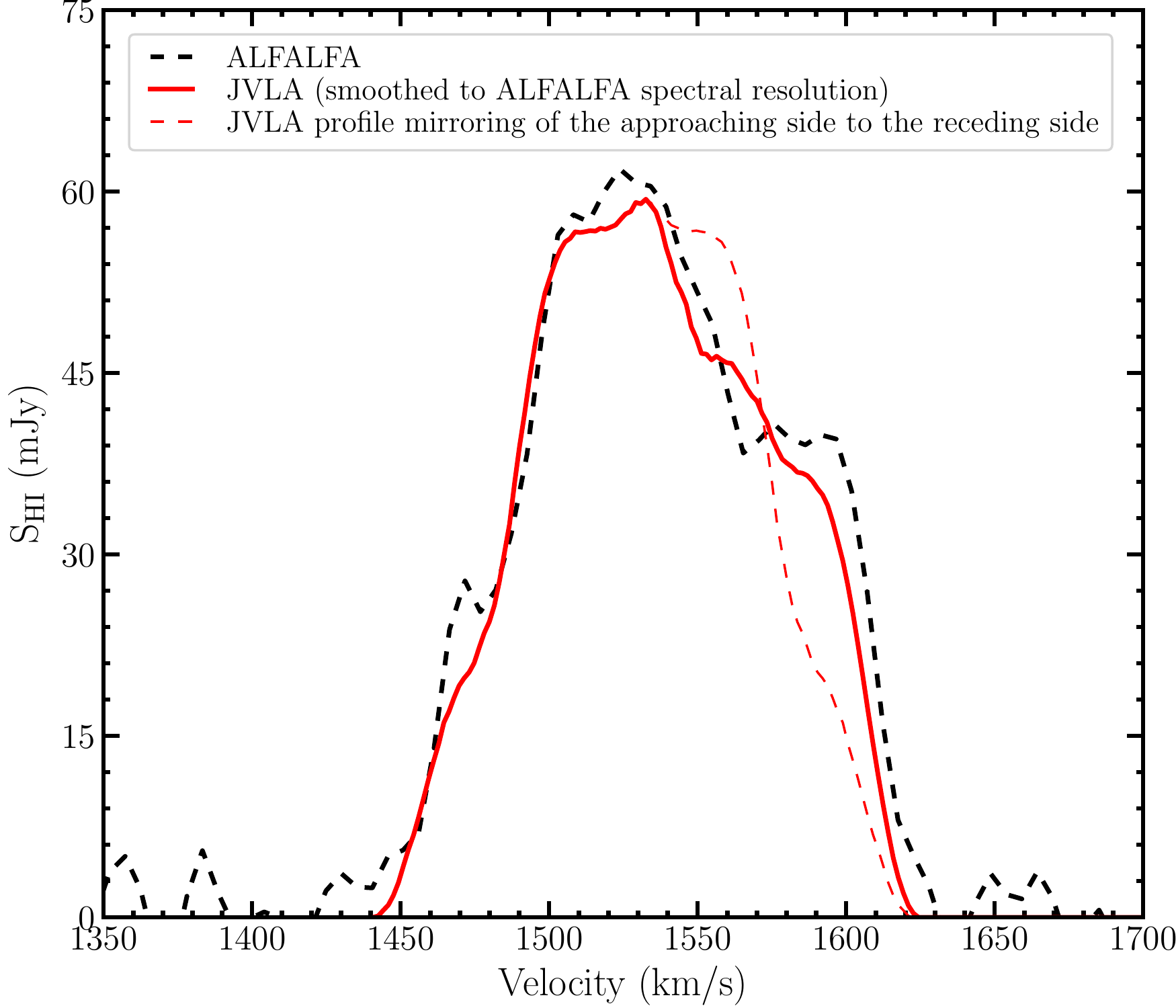}
\caption{
Integrated \HI~line profile of VCC 848.\ Results from ALFALFA and our JVLA observations are plotted as black dashed curve and red solid curve, 
respectively.\ The red dashed curve represents the mirroring of the JVLA profile from the approaching side ($<$ 1532 km s$^{-1}$) to the receding side.\ 
The JVLA \HI~line profile has been smoothed to the 11 km s$^{-1}$ spectral resolution of ALFALFA.\ 
\label{fig_inthispec}}
\end{figure}

The spatially integrated JVLA \HI~line profile of VCC 848 is plotted together with that from the single-dish Arecibo Legacy Fast ALfa Survey (ALFALFA; 
\citealt{haynes11}) observation in Figure \ref{fig_inthispec}.\ The JVLA line profile has been smoothed to the 11 km s$^{-1}$ spectral resolution of ALFALFA 
data.\ The rms noise $\sigma_{\rm rms}$ of the ALFALFA spectrum is about 2.6 mJy.\ The two profiles are in good agreement, except that the JVLA line 
profile is slightly ($<$ 2$\sigma_{\rm rms}$) lower than the ALFALFA profile on the high-velocity side.\ It is worth noting that the JVLA \HI~detection covers 
virtually the same velocity range as the ALFALFA detection.\ The total \HI~flux density measured on the JVLA line profile is 6.30 $\pm$ 0.06 Jy km s$^{-1}$, 
accounting for $\sim$ 94\% of that (6.68 $\pm$ 0.09 Jy km s$^{-1}$) from the ALFALFA observation.\

\section{\HI~Gas Distributions}\label{sec:higas}

\begin{figure*}[ptb]
\centering
\includegraphics[width=1\textwidth]{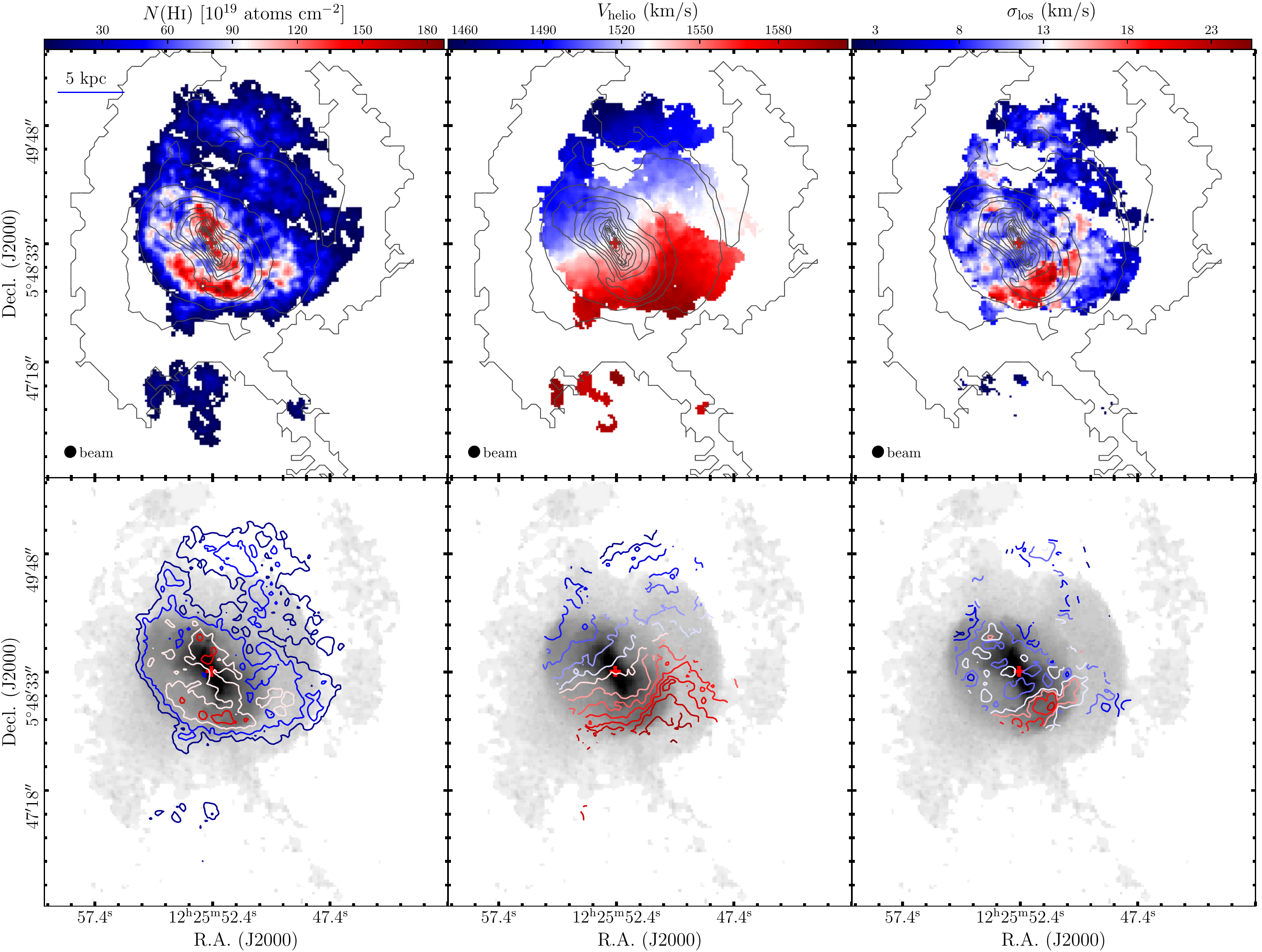}
\caption{
\HI~gas column density map ({\it left}), intensity-weighted velocity field ({\it middle}) and velocity dispersion map ({\it right}) of VCC 848.\
The $g$-band image is contoured onto the \HI~distributions in the top panels, while the \HI~distributions are contoured onto the $g$-band 
image in the bottom panels.\ The $g$-band contour levels run from 19.8 to 25.3 mag arcsec$^{-2}$ with a 0.5 mag interval, and from 25.3 
to 28.3 mag arcsec$^{-2}$ with a 1.0 mag interval.\ Only pixels with velocity-integrated intensities $\geq$ 3 times the rms noise 
are displayed in the velocity field and $\geq$ 5 times the rms noise in the velocity dispersion map.\ The \HI~column density contours are 
at levels of (2, 5, 10, 15) $\times$ 10$^{20}$ cm$^{-2}$.\ The velocity field contours are drawn with 10 km s$^{-1}$ intervals, and the systemic 
velocity of 1532 km s$^{-1}$ is drawn as the yellowish contour lines.\ The red plus symbol in each panel marks the photometric center of the 
system.\ The beam size (7\farcs1 $\times$ 6\farcs6, PA = 56.78$^{\circ}$) is indicated in the bottom-left of the top panels.
\label{fig_himom}}
\end{figure*}

In contrast to the stellar component, the gaseous component of galaxies is collisional and can dissipate its orbital energy via shocks.\ This may lead to diverged 
dynamical evolution of gas and stars during a violent merging process.\ The \HI~21-cm emission has proven to provide indispensable clues to the evolutionary 
status of merging systems \citep[e.g.][]{hibbard01}.\ 

\subsection{Integrated \HI~Line Profile}
The ALFALFA \HI~velocity profile of VCC 848 has a $\sim$ 160 km s$^{-1}$ velocity width\footnote{It is not straightforward to apply an inclination correction to this 
observed velocity width, as we will show that the \HI~gas in VCC 848 appears not to have a simple planar spatial distribution.} measured at 20\% of the peak intensity 
$W_{20}$ \citep{haynes11}.\ This velocity width is larger than that ($\sim$ 130 km s$^{-1}$) expected for its baryonic mass (approximated as 
$M_{\star}+$1.36\MHI~$=$ 7.4$\times$10$^{8}$ $M_{\odot}$), based on the baryonic Tully-Fisher relation \citep[e.g.,][]{mcgaugh12}.\ The \HI~velocity profile is asymmetric 
with respect to the systemic velocity (1532 km s$^{-1}$).\ As we will show later, the approaching half ($V < 1532$ km s$^{-1}$) of the \HI~distribution follows a more regular 
rotation pattern than the receding half ($V > 1532$ km s$^{-1}$).\ So we mirror the \HI~profile of the approaching half to the receding side, and compare the mirrored profile 
with the observed profile at the receding side in Figure \ref{fig_inthispec}.\ It can be seen that the receding side shows a deficit of \HI~at $V \lesssim 1570$ km s$^{-1}$ 
whereas an excess of \HI~at $V \gtrsim 1570$ km s$^{-1}$, indicating significant tidal forces that disturbed the gas velocity distribution.\ $W_{20}$ measured on the mirrored 
line profile is $\simeq$ 140 km s$^{-1}$, which is close to that predicted by the Tully-Fisher relation.\ 

\subsection{\HI~Maps and Comparison with the Stellar Light Distribution}
Figure \ref{fig_himom} shows the column density map, intensity-weighted velocity field and velocity dispersion maps of VCC 848.\ To help visualise the spatial association 
between stellar light and \HI~gas, the $g$-band surface brightness contours are overlaid on the \HI~maps in the upper row, while the \HI~distributions are contoured on 
the $g$-band image in the lower row.\ The $g$ band image has been adaptively smoothed to achieve a minimum signal-to-noise ratio of 5 pixel$^{-1}$ (Paper I).\

The \HI~gas spatial distribution is asymmetric with respect to the photometric center of the stellar main body, and it appears to be confined within the inner edges 
of the two outermost stellar shells.\ Moreover, the \HI~distribution is bounded on the southeast by a near-circular outer edge of the stellar tidal arm that wraps nearly 
180$\degree$ around the galaxy center (Figure 2 in Paper I).\footnote{The spatial association of atomic gas with the stream-like feature reported in Paper I suggests 
that this feature is a tidal arm emanating from the primary progenitor rather than a stellar shell or stream stripped from the secondary progenitor, because the \HI~gas 
of the secondary, if any, should have been completely detached from the secondary's stellar component and assimilated by the primary after the first two to three periapsis 
passages, as will be demonstrated by computer modeling in Section \ref{sec:simul}.}\ Clear-cut edges of the \HI~distribution in the southeastern half are manifested 
by the closely crowded \HI~intensity contour lines.\ \HI~gas distribution in the northwestern half of VCC 848 extends slightly further north than the detected stellar light 
distribution and has typical \NHI~$\lesssim$ 10$^{20}$ cm$^{-2}$ (0.8 $M_{\odot}$ pc$^{-2}$).\ High column density \HI~gas (\NHI~$\gtrsim$ 10$^{21}$ cm$^{-2}$), 
corresponding to the yellowish contours in the lower left panel of Figure \ref{fig_himom}), is mainly concentrated in two distinct spatial areas.\ One is broadly associated 
with the central (twisted) stellar disk of VCC 848 and accounts for $\simeq$ 15\% of the total \HI~flux of the system, while the other one is associated with the above 
mentioned stellar tidal arm and accounts for $\simeq$ 13\% of the total \HI~flux.\ We point out that the general association between the \HI~and stellar features, together 
with the gas-dominated nature of VCC 848, suggest that ram pressure stripping, if any, does not have an appreciable influence on the \HI~morphology.\

The overall \HI~velocity gradient is approximately along the direction of the major axis of the stellar main body, but there exist remarkable local deviations from this overall 
trend.\ In particular, while the velocity field at the approaching side ($V_{\rm los}$ $<$ 1532 km s$^{-1}$) is largely consistent with a solid-body rotation (see below), velocity 
contours of the receding side are closely crowded (i.e.\ steep velocity gradient) and twisted towards the southwest edge of the stellar main body, giving rise to the apparent 
high line-of-sight velocity dispersion (up to $\sim$ 25 km s$^{-1}$) of the \HI~gas there.\ The velocity contours at the receding half is consistent with being affected by tidal 
forces largely along the plane of sky.\ The above mentioned high column density \HI~associated with the tidal arm is mainly at the receding side, where the arm appears to 
be launched from.\ 

By performing multi-gaussian line profile fitting to the \HI~data cube of VCC 848 (see Section \ref{sec:rotcurv}), we find that, except for sporadic locations with high column 
densities near the central disk region, nearly all the other spatial locations have \HI~line profiles that each are adequately fitted by a single Gaussian component.\ This suggests 
that the intensity-weighted velocity field shown in Figure \ref{fig_himom} is largely a reasonable representation of the large scale motion (along the line of sight) of the \HI~gas.\ 
The \HI~channel maps (binned by a factor of 3 along the velocity dimension) are presented in Figure \ref{fig_hichan} of the Appendix.\

\section{Rotation Curves and Mass Profiles}
\label{sec:rotcurv}

\subsection{Derivation of Rotation Curves}
We use the 2D Bayesian Automated Tilted-ring fitter ({\sc 2dbat}) recently developed by \cite{oh18} to extract rotation curves from the \HI~velocity field.\
Instead of using the first moment map which could be affected by the disturbed motions from the merging process, we extract a bulk velocity field for the underlying 
circular rotation of the galaxy in an iterative manner.\ In particular, we perform profile decomposition analysis of the individual velocity profiles of the \HI~data cubes 
as described in \cite{oh19} and extract the decomposed components whose central velocities are close to a reference velocity field (e.g., single Gaussian velocity 
field).\ We call this a bulk velocity field, and classify the other ones deviating from the bulk motions as disturbed, non-circular motions.\ We then improve the 
reference velocity field for profile decomposition with a new model velocity field from the {\sc 2dbat} analysis of the first reference velocity field and repeat the profile 
decomposition described above.\ In this way, we extract the bulk velocity field of VCC 848 for the rotation curve analysis.\ Next, we create a mask for the bulk velocity 
field to isolate a more or less elliptical region encompassing the main body of VCC 848 (left panel of Figure \ref{fig_rotcurv}).\ This quasi-elliptical region has a semi-major 
axis length of $R_{\rm maj}$ $\simeq$ 30\arcsec~and accounts for $\sim$ 30\% of the total \HI~flux of the system.\ The approaching half of this region largely follows the 
characteristic spider-shaped velocity field expected for an inclined rotating disk.\

We fit the velocity field of the masked quasi-elliptical region with {\sc 2dbat} by fixing the center position to the one obtained from our stellar isophotal analysis 
(Table \ref{propertysummary_tab}; Paper I).\ The systemic velocity $V_{\rm sys}$, kinematic PA and inclination angle are fitted as free parameters but are 
kept constant with radius.\ {\sc 2dbat} iteratively searches for the tilted-ring parameters that give the best fit to the observed velocity field.\ The rotation 
velocities (filled circles in the right panel of Figure \ref{fig_rotcurv}) are extracted with a 7\farcs5 ring width that is chosen to be slightly larger than the beam size.\ 
The best-fit kinematic parameters are indicated in the caption of Figure \ref{fig_rotcurv}.\ Lastly, we apply asymmetric drift correction to the extracted rotation 
velocities, following the method described in \cite{bureau02} (see also \citealt{oh11}).\ As is shown in Figure \ref{fig_rotcurv}, the asymmetric drift correction is 
$<$ 2 km s$^{-1}$ across the probed radial range.\

\subsection{Rotation Curves}\label{sec:rotcurvslope}
Rotation velocities extracted from the approaching side increase nearly linearly with radius at $R_{\rm maj}$ $\lesssim$ 20\arcsec, beyond which the rotation curve 
appears to flatten out.\ At the receding side, where the velocity field is substantially disturbed by the ongoing merging, the derived circular velocities are about half 
that of the approaching side at $R_{\rm maj}$ $\lesssim$ 25\arcsec, beyond which the radial trend steepens and reaches velocities comparable to that of the approaching 
side at $R_{\rm maj}$ $\sim$ 30\arcsec.\ Projected radial velocities derived based on the extracted rotation velocities are over-plotted on the \HI~position-velocity diagrams 
in Figure \ref{fig_pv}, where the largely regular rotation pattern in the inner 30\arcsec~is clearly illustrated.\

\begin{figure*}[ptb]
\centering
\includegraphics[width=1\textwidth]{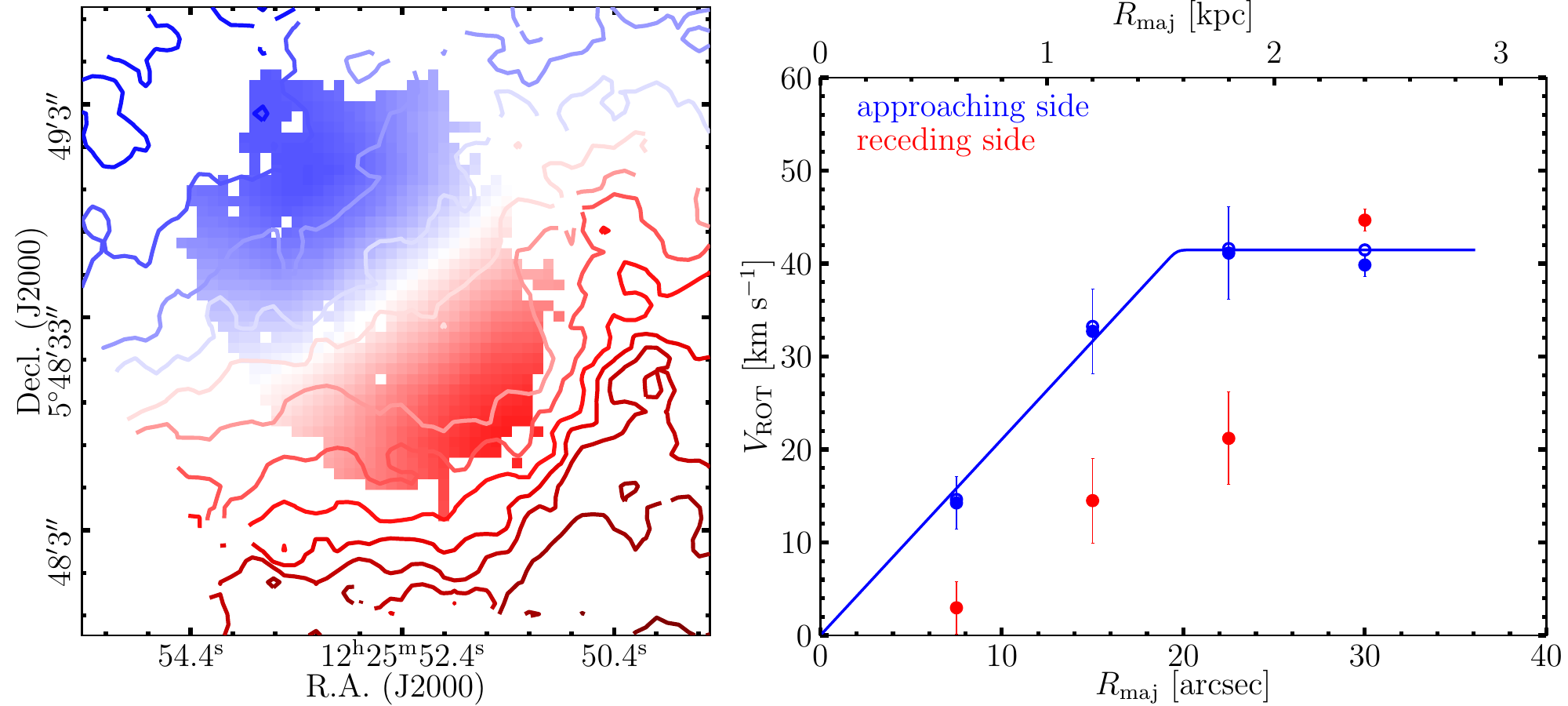}
\caption{
Results of the \HI~velocity field modeling with the {\sc 2dbat} algorithm.\
The {\it left} panel shows the model velocity field constructed using the best-fit tilted-ring parameters, where $V_{\rm sys}$ = 1532 km s$^{-1}$, 
PA = 21.8\degree~and inclination angle = 45.2\degree.\ The same velocity contours shown in Figure \ref{fig_himom} for the observed velocity 
field are overlaid on the model velocity field.\ The {\it right} panel shows the rotation velocities determined for the approaching ({\it blue}) and receding 
({\it red}) sides separately.\ Note that the receding side (to the southwest) is severely distorted by the ongoing merging event.\ The approaching side 
is largely consistent with a solid-body rotation that can be reasonably modeled.\ The filled symbols represent the extracted rotation velocities, and the 
open circular symbols represent the asymmetric drift corrected rotation velocities.\ The {\it blue} curve represents a parametric fitting to the rotation 
curve of the approaching side by adopting the same functional form used in \cite{berrera18}.\ See Section \ref{sec:rotcurvslope} for details.\
\label{fig_rotcurv}}
\end{figure*}

\begin{figure*}[ptb]
\centering
\includegraphics[width=1\textwidth]{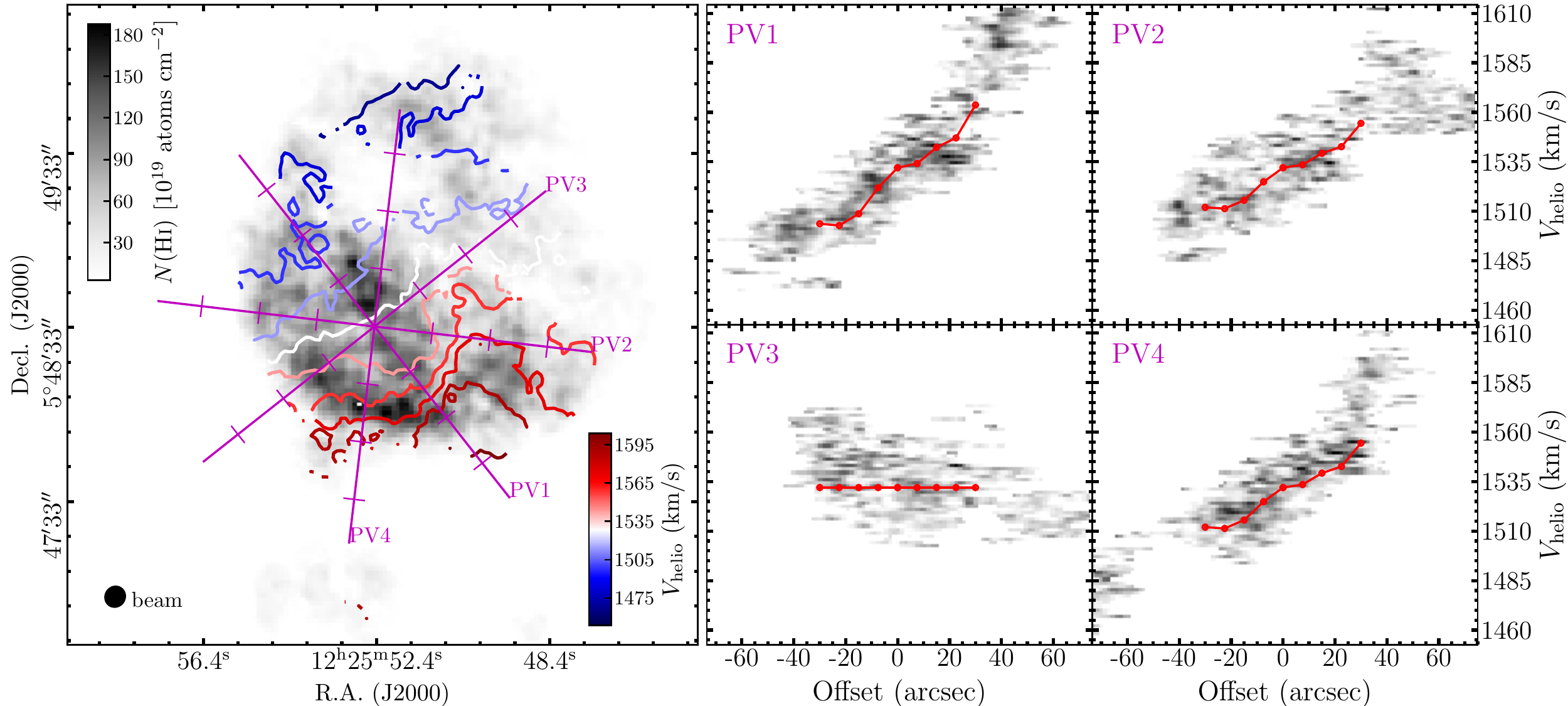}
\caption{
\HI~position-velocity diagrams ({\it right} panels) extracted along four 7\arcsec-wide pseudo-slits oriented at PA $=$ 38.3\degree (PV1), 
83.3\degree (PV2), 128.3\degree (PV3) and 173.3\degree (PV4), respectively, where 38.3\degree~is the PA of the best-fit kinematic major 
axis of VCC 848 (Figure \ref{fig_rotcurv}).\ The four pseudo-slits (centered at the galaxy center) are indicated as long magenta lines overlaid 
on the \HI~column density map ({\it left} panel).\ Every 20\arcsec~along the pseudo-slit direction is marked with a 7\arcsec-long line perpendicular 
to the slit direction.\ Also overlaid on the column density map are the intensity-weighted \HI~velocity contours, with 15 km s$^{-1}$ intervals.\ 
In the {\it right} panels, the dot-connected red curves are the projected radial velocities based on rotation curves determined for the central 
30\arcsec~of VCC 848 (Figure \ref{fig_rotcurv}).
\label{fig_pv}}
\end{figure*}

In order to quantify the radial trend of rotation velocities extracted from the approaching side, we parameterize the rotation curve with the same functional 
form as that used in \cite{berrera18} (see their Equation 4).\ The functional form is defined by three parameters: the maximum velocity ($V_{\rm max}$), the 
transition radius ($R_{\rm turn}$) where the rotation curve transitions from a solid-body inner part to a flat outer part and the sharpness ($\alpha$) of the transition.\ 
We obtain a best-fit $V_{\rm max}$ $=$ 41.5$\pm$0.5 km s$^{-1}$, $R_{\rm turn}$ $=$ 19.6\arcsec$\pm$0.9\arcsec~($\sim$ 1.6 kpc) and $\alpha$ $=$ 130.1$\pm$1.0.\
The large $\alpha$ value indicates a sharp transition at $R_{\rm turn}$.\ The corresponding best-fit curve is over-plotted in the right panel of Figure \ref{fig_rotcurv}.\

Based on the above parametrized fitting, we obtain a slope of 26.5$\pm$3.9 km s$^{-1}$ kpc$^{-1}$ ($V_{\rm max}$/$R_{\rm turn}$) for the solid-body rising part 
of the rotation curve, where the uncertainty is determined based on repeat fitting to randomly disturbed rotation velocities for 1000 times.\ This slope of rotation curve 
falls within the range of ordinary dwarf irregular galaxies or BCDs with off-centered starburst that have inclination-corrected central brightness ($\mu_{g, {\rm corr}}$ 
= 22.6 mag arcsec$^{-2}$ for a kinematic inclination $=$ 45.2\degree) comparable to VCC 848 \citep[][]{lelli14}.\ Because the rotation velocity gradient slope is 
proportional to the square root of local matter density for a solid-body rotation curve, we can infer that the merging event of VCC 848 has not built up an exceptionally 
compact core.\ 

\subsection{Dynamical and Baryonic Mass Profiles}\label{sec:dynbaryprofiles}
\begin{figure}[ptb]
\centering
\includegraphics[width=0.47\textwidth]{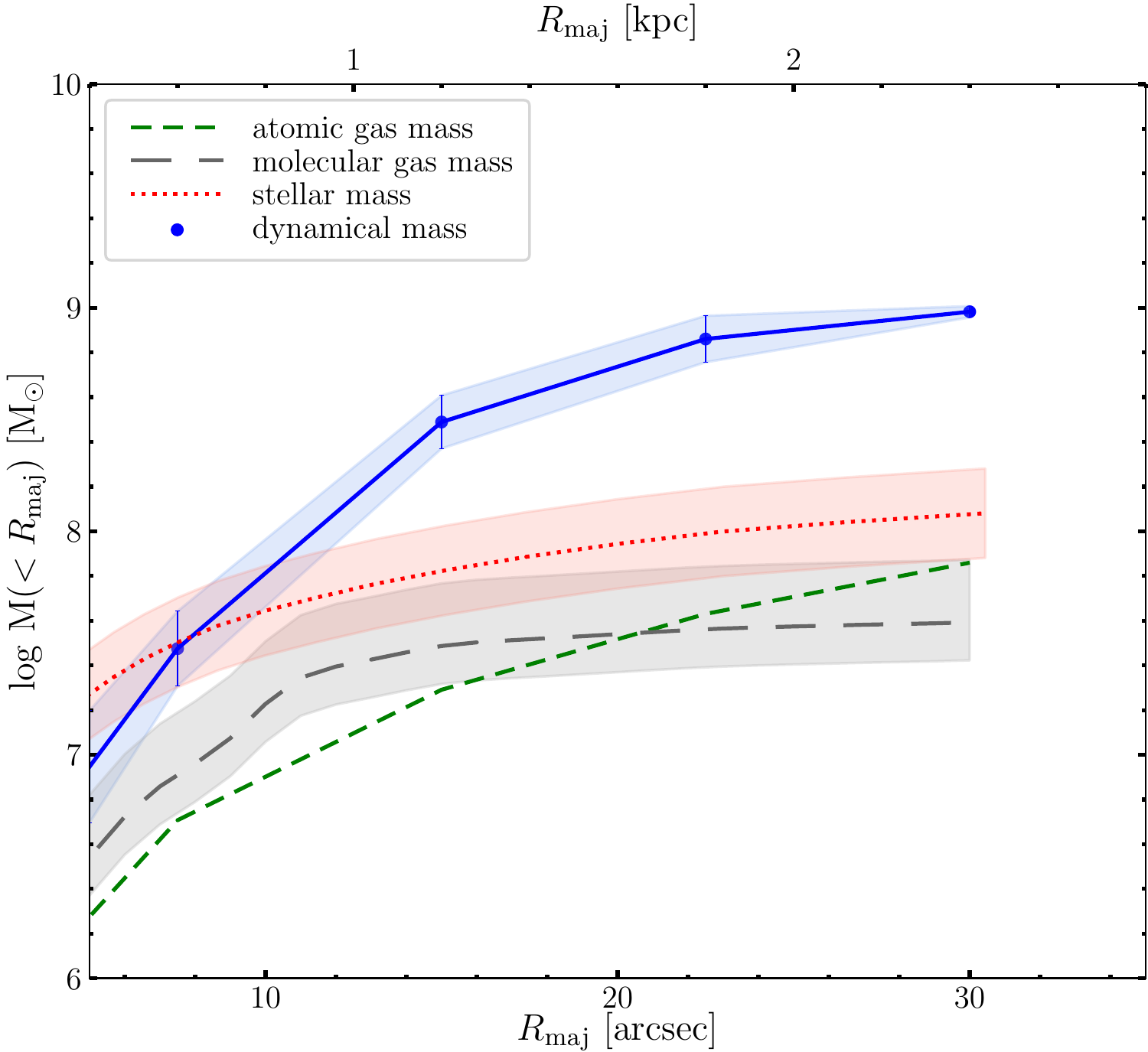}
\caption{
Accumulated radial profiles of the dynamical mass as well as the stellar, atomic and molecular gas masses within the central 30\arcsec.\
The same geometric parameters derived from the {\sc 2dbat} fitting of the \HI~velocity field (Figure \ref{fig_rotcurv}) are used for extracting all  
of the profiles.\ The atomic gas mass is calculated as the \HI~gas mass multiplied by a factor of 1.36 to account for the contribution of helium.\ 
Estimation of the molecular gas mass is described in Section \ref{sec:dynbaryprofiles}\ The gray-shaded region represents the uncertainties of 
the estimation of molecular gas mass.\
\label{fig_massprofile}}
\end{figure}

We use the rotation curve of the approaching half to determine the cumulative dynamical mass profiles (= $V_{\rm rot}(R)^{2}$$R$/G), which is shown in 
Figure \ref{fig_massprofile}.\ We also overplot mass profiles of the baryonic components, including stars, atomic and molecular gas in Figure \ref{fig_massprofile}.\ 
All of the profiles are extracted using the same geometric parameters as determined from the {\sc 2dbat} kinematic fitting.\ The stellar mass is estimated by using 
the $(g-i)-$mass-to-light relation calibrated for Local Group dwarf galaxies \citep{zhang17}.\ Details about the procedure of stellar mass estimation are given in 
Paper I.\ We adopt a logarithmic stellar mass uncertainty of 0.2 dex which is applicable to color-based mass estimation for low-dust-extinction cases \citep{zhang17}.\ 
The atomic gas mass is equal to the \HI~gas mass multiplied by a factor of 1.36 to account for helium.\ The molecular gas mass is indirectly inferred from the SFR 
estimated with H$\alpha$ luminosities \citep[][Equation 10]{catalan15}, by adopting a constant molecular gas consumption timescale of $(5.25\pm2.5)\times10^{-10}$ 
yr$^{-1}$ found for nearby disk galaxies at (sub-)kpc scales \citep{bigiel08,leroy08}.\footnote{Another method of estimating the molecular gas mass is based on a 
metallicity-dependent gas-to-dust ratio (G/D).\ According to the G/D$-$metallicity relation from \cite{remy14}, VCC 848 is expected to have a $\log$(G/D) of 3.00 
$\pm$ 0.37 for 12$+\log$(O/H) $=$ 8.03 (Table \ref{propertysummary_tab}).\ In addition, \cite{grossi15} determined a total dust mass of 10$^{5.01}$ $M_{\odot}$ 
(the value has been adjusted to reflect our adopted distance) within the central $\sim$ 30\arcsec~of VCC 848.\ Therefore, the total gas mass (atomic plus molecular) 
within the central 30\arcsec~is estimated to be 10$^{8.01 \pm 0.37}$ $M_{\odot}$ and hence the most likely molecular gas mass be $\sim$ 10$^{7.5}$ $M_{\odot}$ 
by subtracting the atomic gas contribution (Figure \ref{fig_massprofile}).\ This molecular gas mass is in reasonable agreement with that (10$^{7.6}$ $M_{\odot}$) inferred 
from the SFR-based method shown in Figure \ref{fig_massprofile}.\ We choose to use the SFR-based method in this paper.\ This is because, besides their low spatial 
resolution, the {\it Herschel} images of VCC 848 generally have signal-to-noise ratios that are too low (except for the brightest star-forming sites) to allow for constructing 
a robust radial profile of dust mass densities.}

As is illustrated in Figure \ref{fig_massprofile}, the stellar and atomic gas components together account for $\simeq$ 20\% of the dynamical mass enclosed 
within the central 30\arcsec, and the fraction increases to $\sim$ 24\% if including molecular gas in the baryon budget.\ 
Stars constitute more than half (54\%) of the baryonic mass, which is in contrast to the overwhelmingly gas-rich nature of the system as a whole.\ This might 
be partly attributed to a highly efficient star formation toward the stellar main body and a less severe tidal stripping of the stellar component than the gas during 
the merging.\ Lastly, it is also notable that the molecular gas appears to dominate over the atomic gas in mass within the central $\sim$16\arcsec$-$20\arcsec 
(1.3$-$1.6 kpc; accounting for the uncertainties of molecular gas mass estimate) in radius.\ Similarly high fraction of molecular gas was also indirectly inferred for 
some nearby BCD or starburst dwarf galaxies \citep{hunter19}.\

\section{Star Formation Activities}
\subsection{Specific SFR and Star Formation Efficiencies}\label{sec:ssfrsfe}
\label{sec:sfrsfe}
\begin{figure*}[ptb]
\centering
\includegraphics[width=1\textwidth]{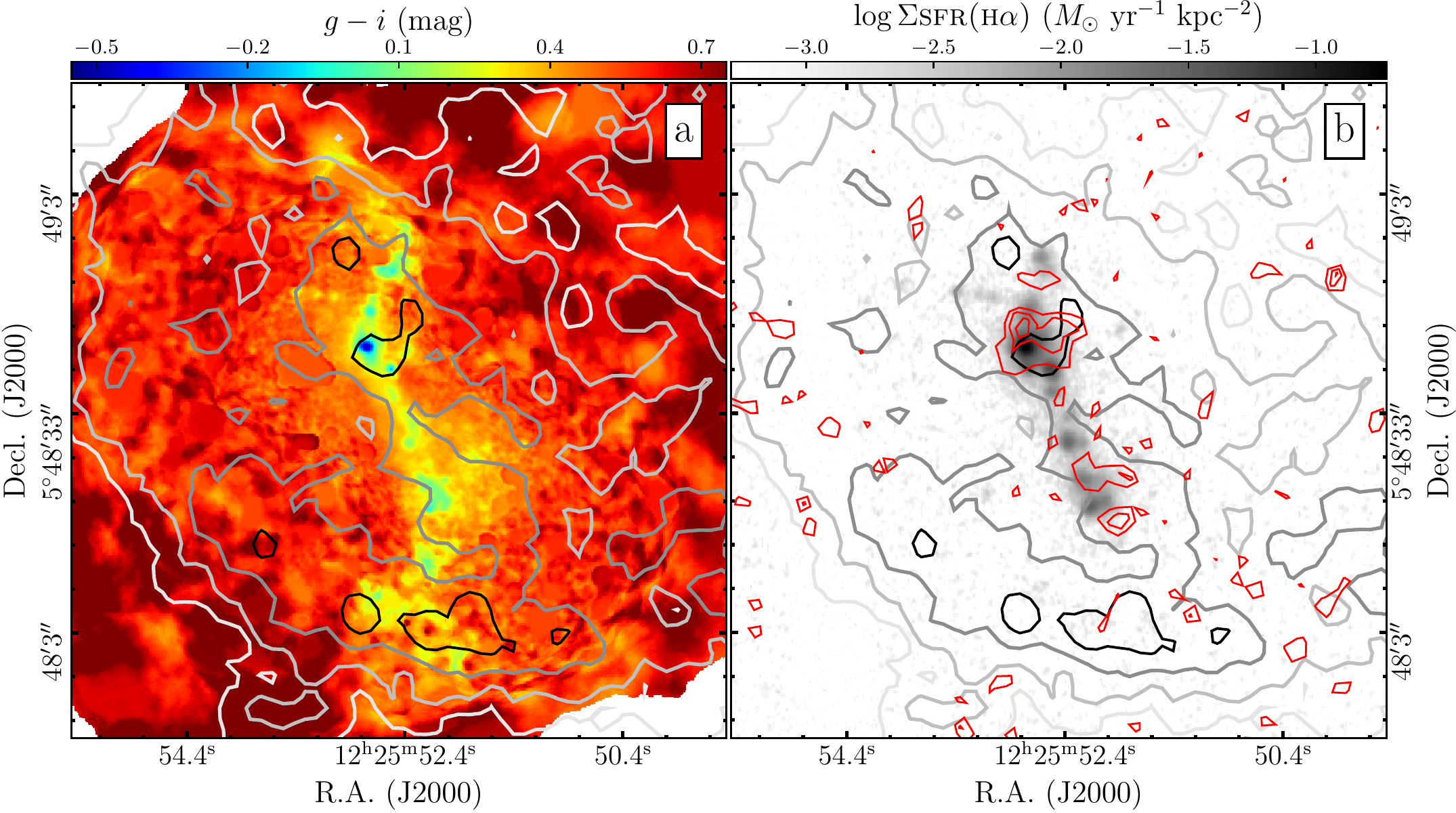}
\caption{
$g-i$ color map ({\it left}) and H$\alpha$ flux density map ({\it right}) of the central 1.5$\times$1.5 arcmin of VCC 848.\ 
The color bar for the H$\alpha$ map is labeled with the corresponding SFR surface density, by adopting the recipe from \cite{catalan15}.\ 
The same \HI~column density contours shown in Figure \ref{fig_himom} are overlaid in both panels here.\ The {\it Herschel} 100$\mu$m 
is overlaid as red contours (2$\sigma$, 3$\sigma$ and 4$\sigma$ above the sky background) on the H$\alpha$ flux  density map.\ 
Note that the dust IR emission contributes at most $\sim$ 10\% of the total SFR budget (H$\alpha$+TIR) across the system at spatial scales 
comparable to the resolution (9$\farcs$4) of the 100 $\mu$m image.
\label{fig_clrsfr}}
\end{figure*}

The global logarithmic specific SFR (SFR per unit stellar mass, sSFR) of VCC 848 is $-9.87$ (see Table \ref{propertysummary_tab} for the total SFR and 
stellar mass), which is 0.19 dex higher than that expected for normal star-forming galaxies of similar stellar masses \citep[Equation 19]{shin19}.\ However, 
given a 0.27 dex scatter of the \cite{shin19} relation, VCC 848 is consistent with being a main sequence star-forming galaxy for its stellar mass.\ The 
($g-i$) and H$\alpha$ maps of the central 1.5$\times$1.5 arcmin are shown in Figure \ref{fig_clrsfr}.\ The {\it Herschel} 100 $\mu$m continuum emission, 
which traces the dust obscured star formation, is contoured in red color on the H$\alpha$ map.\ A vast majority ($\simeq$ 90\%) of current star formation 
traced by H$\alpha$ and 100 $\mu$m is confined to the central high surface brightness regions enclosed by the \NHI~$=$ 10$^{21}$ cm$^{-2}$ contour.\ 
Although the tidal arm in the southeast reaches similar \NHI~to the central region, the SFR there accounts for merely $\simeq$ 6\% of the total.\ The most 
intense star-forming site of the system is located $\sim$ 10$\arcsec$ offset toward the northeast of the photometric center, and is associated with the brightest 
star cluster (Paper I) and \HII~region (Section \ref{sec:hiiregions}) in VCC 848.\

It is remarkable that the youngest stellar populations in the central $\sim$ 10\arcsec~(in radius) are confined to a straight narrow strip that appears to trace the 
symmetry axis of the twisted disk.\ The age spread of star clusters detected along this narrow strip suggest that it has existed for at least $\sim$ 1 Gyr (Paper I).\ 
We note that this linear stellar structure is not likely a bar, because the associated \HI~velocity field does not show the characteristic S-shaped isovelocity contours 
expected for a bar structure \citep[e.g.,][]{athanassoula92}.\

\begin{figure*}[ptb]
\centering
\includegraphics[width=1\textwidth]{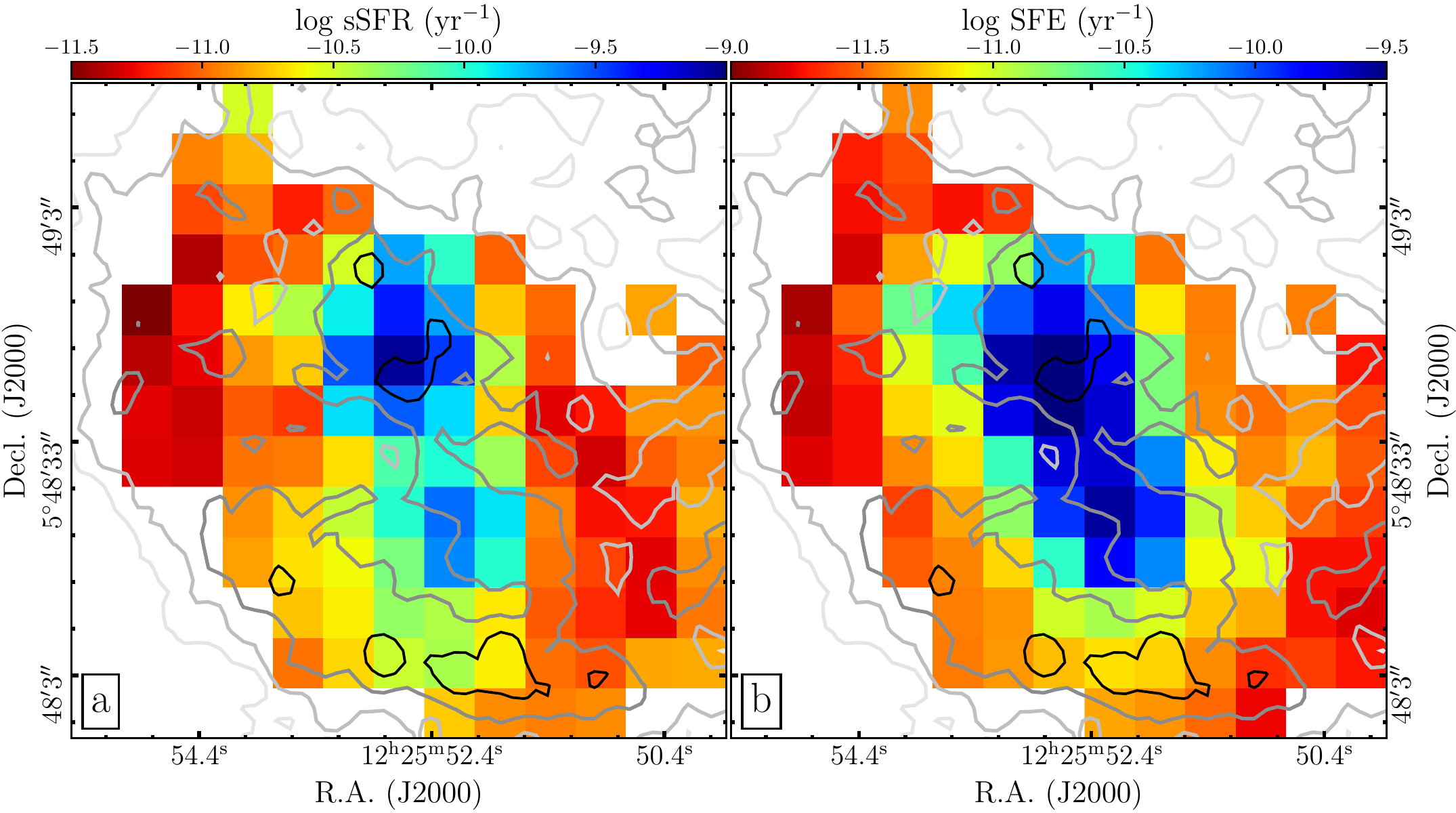}
\caption{
Maps of specific SFR (sSFR, {\it left}) and star formation efficiency (SFE) with respect to atomic plus molecular gas ({\it right}) of the 
central 1.5$\times$1.5 arcmin of VCC 848.\ As in Figure \ref{fig_clrsfr}, the SFR is estimated based on the observed H$\alpha$ flux intensities.\ 
The H$\alpha$ and optical images have been smoothed to match the spatial resolution of \HI~data and re-binned to a 7$\arcsec$$\times$7$\arcsec$ 
pixel size before deriving the maps shown here.\ The same \HI~column density contours shown in Figure \ref{fig_himom} are overlaid in both panels.\ 
See Section \ref{sec:sfrsfe} for details.
\label{fig_ssfrsfe}}
\end{figure*}

To further illustrate the relative strength of star formation activities across the system, we show spatial distributions of sSFR and star formation 
efficiency (SFE) with respect to the atomic plus molecular gas (see Section \ref{sec:rotcurv} for the inference of molecular gas mass) in 
Figure \ref{fig_ssfrsfe}, where the involved input images, including the $g$, $i$, H$\alpha$ and \HI, have been smoothed to match the beam size 
of \HI~map and rebinned to 7$\arcsec$$\times$7$\arcsec$ (560$\times$560 pc) pixel sizes.\ The SFR is estimated from the H$\alpha$ luminosities.\ 
Note that we have ignored the contribution of dust obscured star formation in Figure \ref{fig_ssfrsfe} for a high resolution view of the overall distribution 
of star formation.\ This practice is justified by the fact that the TIR inferred from the 100 $\mu$m (\citealt{galametz13}) accounts for at most $\sim$ 10\% 
of the local SFR at spatial scales comparable to the resolution of 100 $\mu$m, according to the hybrid H$\alpha$+FIR recipe for SFR estimation presented 
in \cite{catalan15}.\ This is probably not surprising given the low metallicities of VCC 848.\ 

\begin{figure}[ptb]
\centering
\includegraphics[width=0.47\textwidth]{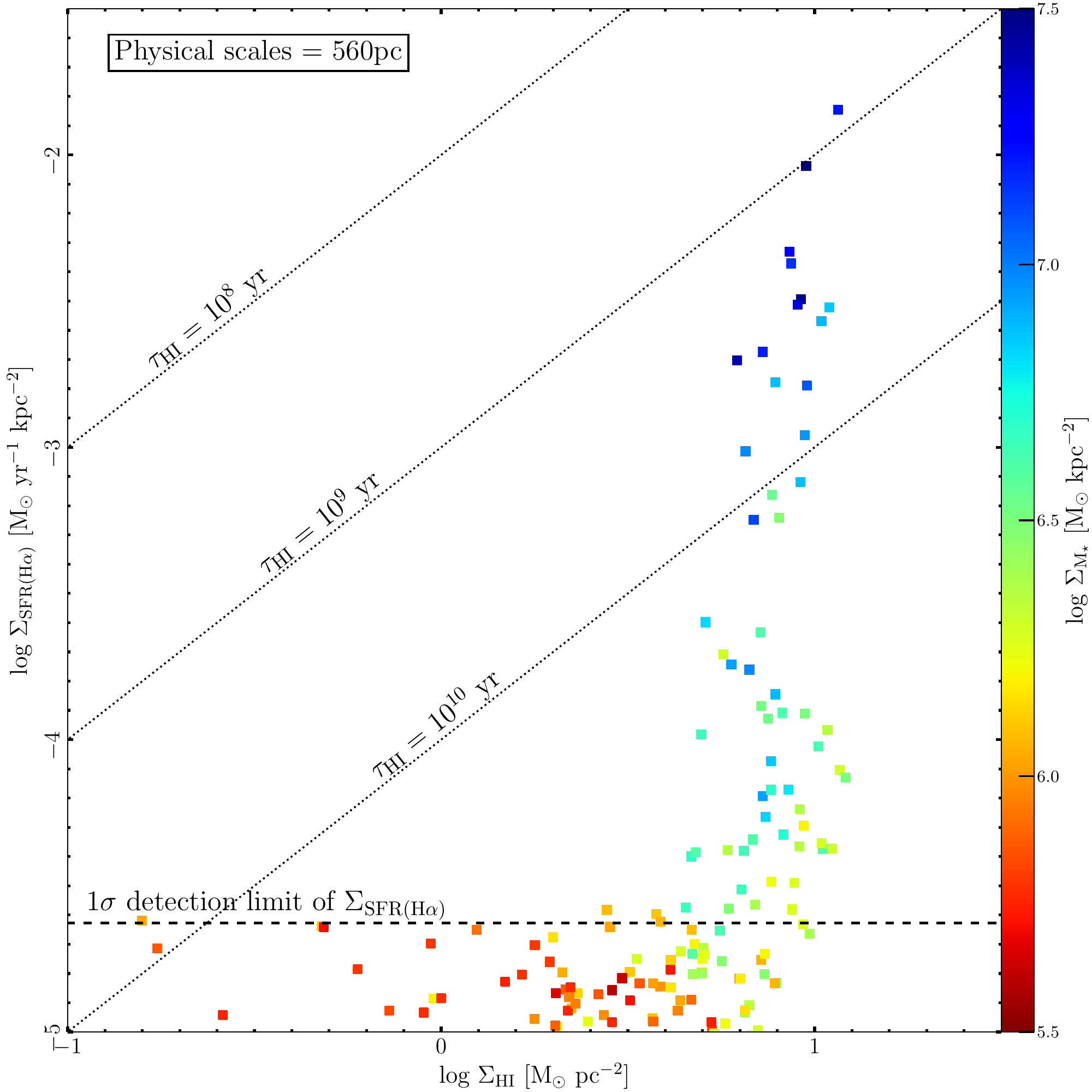}
\caption{
$\Sigma_{\rm SFR(H\alpha)}$ is plotted as a function of \SIGMAHI~at 560 pc resolution.\ The data points correspond to the individual pixels plotted 
in Figures \ref{fig_clrsfr} and \ref{fig_ssfrsfe}, and are color-coded according to their stellar mass surface densities.\ Note that only data points that fall 
within the 5$\times$10$^{20}$ cm$^{-2}$ contours (Figure \ref{fig_himom}) are plotted.\ The diagonal dotted lines represent lines of constant \SFEHI~and 
thus constant \HI~gas depletion (by star formation after transitioning to the molecular phase) times \TAUHI~(10$^{8}$, 10$^{9}$ and 10$^{10}$ yr).\ 
Above the detection limit of SFR(H$\alpha$), the data points fall into two groups which are separated by a 0.3 dex gap in log $\Sigma_{\rm SFR(H\alpha)}$ 
from $-3.6$ to $-3.3$.\ Data points of the group with higher log $\Sigma_{\rm SFR(H\alpha)}$ belong to the central high surface brightness star-forming regions, 
whereas those of the lower-$\Sigma_{\rm SFR(H\alpha)}$ group with log \SIGMAHI~$\gtrsim$ 0.88 are from the southeast \HI~tidal arm (Figure \ref{fig_himom}).\
\label{fig_plotsfrhi}}
\end{figure}

As is shown in Figure \ref{fig_ssfrsfe}, the sSFR and SFE reach their logarithmic maximum of $\simeq$ $-9.1$ yr$^{-1}$ and $-9.4$ yr$^{-1}$ respectively 
at the most intense (and also bluest) star-forming site of the system.\ The \HI~tidal arm has typical sSFR and SFE that are more than one order of magnitude 
lower than the central high surface brightness regions.\ In Figure \ref{fig_plotsfrhi}, the SFR surface densities of individual pixels shown in Figure \ref{fig_ssfrsfe} 
are plotted as a function of the \HI~mass surface densities.\ Above a detection limit of $10^{-4.6}$ $M_{\odot}$ yr$^{-1}$ kpc$^{-2}$ for $\Sigma_{\rm SFR(H\alpha)}$, 
the individual regions fall into two groups which are separated by a $\sim$ 0.3 dex gap in $\log\Sigma_{\rm SFR(H\alpha)}$ from $-3.6$ to $-3.3$.\ The regions in the 
higher-$\Sigma_{\rm SFR(H\alpha)}$ group belong to the central high surface brightness regions, whereas those in the lower-$\Sigma_{\rm SFR(H\alpha)}$ group 
and with $\log$\SIGMAHI~$\gtrsim$ 0.88 belong to the \HI~concentration along the tidal arm (Figure \ref{fig_himom}).\ The lack of a direct connection between star 
formation and \HI~gas and the apparent relevance of stellar surface densities are vividly demonstrated in VCC 848 (see, e.g., \citealt{ekta08} for similar examples).\ 

\subsection{H$\alpha$ Luminosity Distribution of \HII~Regions}
\label{sec:hiiregions}

%\Distribution of the best-fit surface brightness profile models.
%\begin{longrotatetable}
\begin{deluxetable}{lcccc}
\tabletypesize{\small}
\tablecolumns{5}
\setlength{\columnsep}{40pt}
\tablewidth{0pt}
\tablecaption{\label{halphalf_tab} H$\alpha$ luminosities of the brightest \HII~regions in VCC 848}
\tablehead{
\colhead{ID}
&\colhead{R.A.}
&\colhead{Decl.}
&\colhead{$\log$$L_{\rm H\alpha}$}
&\colhead{$\sigma$(log$L_{\rm H\alpha}$)} \\
\colhead{}
&\colhead{(degree)}
&\colhead{(degree)}
&\colhead{(erg s$^{-1}$)}
&\colhead{(erg s$^{-1}$)} \\
\colhead{(1)}
&\colhead{(2)}
&\colhead{(3)}
&\colhead{(4)}
&\colhead{(5)} \\
\noalign{\vskip -4.5mm}
}

\startdata
\noalign{\vskip -0.5mm}
%\hline
1  &  186.46985  &  5.81168  &  38.94  &  0.01 \\
2  &  186.46889  &  5.81087  &  38.36  &  0.02 \\
3  &  186.46735  &  5.80558  &  38.16  &  0.02 \\
4  &  186.46820  &  5.80808  &  38.06  &  0.03 \\
5  &  186.46973  &  5.81301  &  37.99  &  0.04 \\
6  &  186.46762  &  5.80660  &  37.96  &  0.04 \\
7  &  186.46974  &  5.81087  &  37.92  &  0.02 \\
8  &  186.46677  &  5.80580  &  37.87  &  0.03 \\
9  &  186.46882  &  5.81001  &  37.82  &  0.03 \\
10  &  186.46925  &  5.81512  &  37.74  &  0.05 \\
%11  &  186.46818  &  5.80905  &  37.46  &  0.10 \\
%12  &  186.47245  &  5.81379  &  37.32  &  0.08 \\
%13  &  186.47108  &  5.81356  &  37.12  &  0.19 \\
%14  &  186.47463  &  5.81406  &  37.08  &  0.13 \\
%15  &  186.47162  &  5.81367  &  37.05  &  0.16 \\
%16  &  186.46622  &  5.81234  &  37.05  &  0.16 \\
%17  &  186.47134  &  5.80608  &  37.03  &  0.15 \\
%18  &  186.46561  &  5.80740  &  36.99  &  0.22 \\
%19  &  186.46624  &  5.80412  &  36.83  &  0.32 \\
%20  &  186.46673  &  5.80120  &  36.79  &  0.29 \\
%21  &  186.46618  &  5.80464  &  36.77  &  0.45 \\
%22  &  186.47594  &  5.81345  &  36.74  &  0.28 \\
%23  &  186.46913  &  5.80266  &  36.59  &  0.48 \\
\enddata
%\tablecomments{
%Col 1: JVLA configuration; Col 2: observing date; Col 3: time on source; Col 4: number of channels; Col 5: channel width in km s$^{-1}$).
%}
%\label{halphalf_tab}

\end{deluxetable}
%\end{longrotatetable}

\subsubsection{Detection of \HII~Regions}
\HII~regions trace massive stars ($\gtrsim$ 10 $M_{\odot}$) formed in the recent ($\sim 10-20$ Myr) past, and their spatial and luminosity distributions reflect the 
mode and strength of star-forming activities.\ We use the {\sc Sextractor} software \citep{bertin96} to detect \HII~regions based on the H$\alpha$ image of VCC 848.\ 
With a detection threshold of 2.5 times the rms noise, we detect 23 \HII~regions in total.\ To determine the detection completeness, we repeatedly add artificial point 
sources (one at a time) that are convolved with a 1.8\arcsec~FWHM Gaussian (Section \ref{sec:hiiphot}) and span a uniform range of flux densities at 1000 randomly 
selected positions within the central main body of VCC 848 (delineated by the dotted ellipse overplotted on the H$\alpha$ image in Figure \ref{fig_plothalf}), and then run 
{\sc Sextractor} for detection with the same parameters as for the original image.\ This test suggests that our \HII~region detection reaches $\geq$ 90\% recovery rate 
(our completeness limit) at $\log L$(H$\alpha$) [erg s$^{-1}$] $\gtrsim$ 37.7.\ Among the 23 \HII~regions detected in VCC 848, 10 have $L$(H$\alpha$) above the 90\% 
limit.\ The following analysis in this section will focus on the 10 brightest \HII~regions.

\subsubsection{Aperture Photometry of \HII~Regions}\label{sec:hiiphot}
An inspection of the radial light distributions of the 10 brightest \HII~regions suggests that they have FWHMs ranging from 1.7\arcsec~to 1.9\arcsec~and nearly all of them 
are reasonably described by single Gaussian profiles\footnote{The similar radial profiles might not be very surprising given that our \HII~regions are only marginally    
resolved in the H$\alpha$ image and that the intrinsic light profiles of \HII~regions are found to have nearly invariant gradients at $\log L$(H$\alpha$) [erg s$^{-1}$] 
$\lesssim$ 38.6 \citep{rozas98}.}.\ 
We perform photometry for each region with a 2.1\arcsec~(8 pixels) diameter circular aperture and subtract a local median background using an 8-pixel wide annulus 
that is sufficiently far away from the center of the \HII~region in question.\ The aperture size is chosen to be larger than the FWHM but at the same time to avoid significant 
contamination from neighboring regions.\ Pixels belonging to neighboring \HII~regions are excluded when determining local background and noise.\ For the two \HII~regions 
close to the brightest \HII~region (left panel of Figure \ref{fig_plothalf}), we manually select background regions that are at the same distance from the brightest \HII~region 
but in different directions.\ The total H$\alpha$ flux of each region, as expected for a Gaussian profile with FWHM of 1.8\arcsec, is obtained by applying a multiplicative 
aperture correction factor of 1.77 to the background-subtracted aperture photometry.\ The derived H$\alpha$ flux densities and uncertainties of the 10 \HII~regions are 
given in Table \ref{halphalf_tab}.\ We note that the marginally-resolved nature of our detections means that more sophisticated procedures for \HII~region photometry 
such as HIIphot \citep{thilker00} cannot play their advantages.\

\subsubsection{\HII~Region Distributions}
The \HII~region distributions are shown in Figure \ref{fig_plothalf}, where the left panel marks the spatial distribution while the right panel presents the cumulative 
H$\alpha$ luminosity distribution above the 90\% completeness limit.\ No binning in luminosity is used for constructing the cumulative distribution.\ The variation of detection 
completeness with luminosities are also overplotted in the right panel.\ To provide some perspective on the \HII~regions detected in VCC 848, the 90\% completeness limit of 
$L$(H$\alpha$) is $\sim$ 5 times that of the Orion nebula \citep{kennicutt84}, and for a Case B recombination, this corresponds to a Lyman continuum photon emission rate 
of 10$^{49.6}$ s$^{-1}$, which is equivalent to $\sim$ 13 O9V stars \citep{sternberg03}.\ The brightest \HII~region is associated with the brightest star cluster in VCC 848 (Paper I) 
and has a H$\alpha$ luminosity (10$^{38.94}$ erg s$^{-1}$) slightly higher than the brightest \HII~region N66 (10$^{38.78}$ erg s$^{-1}$, \citealt{kennicutt84}) in the Small 
Magellanic Cloud (SMC).\ We note that the H$\alpha$ luminosity of the brightest \HII~region may be excited by the associated (and the brightest) star cluster which has an 
age of $\sim$ $6-8$ Myr and mass of $\simeq$ 10$^{5}$ $M_{\odot}$ (Paper I).\

We perform a least-squares fitting to the completeness-corrected cumulative H$\alpha$ luminosity distribution of VCC 848 by adopting a power-law form for the (probability) 
luminosity distribution ($dN(L)/dL\propto L^{-\alpha}$) that has been widely used in the literature.\ We find a best-fit power-law index $\alpha = 2.6^{+0.7}_{-0.5}$, where the 
uncertainties correspond to the range of $\alpha$ with reduced $\chi^{2} \leq \chi_{min}^2+1$.\ As is illustrated in Figure \ref{fig_plothalf}, the luminosity distribution is well 
described by a single power-law form, except for the brightest \HII~region which is $\sim$ 0.2 dex more luminous than that expected for the power-law fitting.

Considering the uncertainties, the power-law index determined above for VCC 848 falls in the broad range of values found for nearby spiral galaxies ($\sim$ $2.0\pm0.5$; 
e.g.\ \citealt{kennicutt89,helmboldt05, liu13}), but appears to be at the upper bound for dIrr galaxies ($\sim$ $1.5\pm0.5$; e.g.\ \citealt{strobel91, miller94, youngblood99, 
vanzee00}).\ However, it has been shown that measurements of $\alpha$ are in many cases dependent on the $L$(H$\alpha$) range used in the power-law fitting, in the 
sense that $\alpha$ measured at higher luminosities tends to be similar to or larger than that at lower luminosities \citep[e.g.,][]{youngblood99}.\ The \HII~region detection 
limit for VCC 848 is relatively high compared to most previous studies of dIrr galaxies in the literature, so the large $\alpha$ measured for VCC 848 may not be surprising.\ 
We shall return to this point when comparing VCC 848 with ordinary dIrr galaxies in the next section.\

\subsubsection{Comparison with Isolated dIrr Galaxies}

\begin{figure*}[ptb]
\centering
\includegraphics[width=1\textwidth]{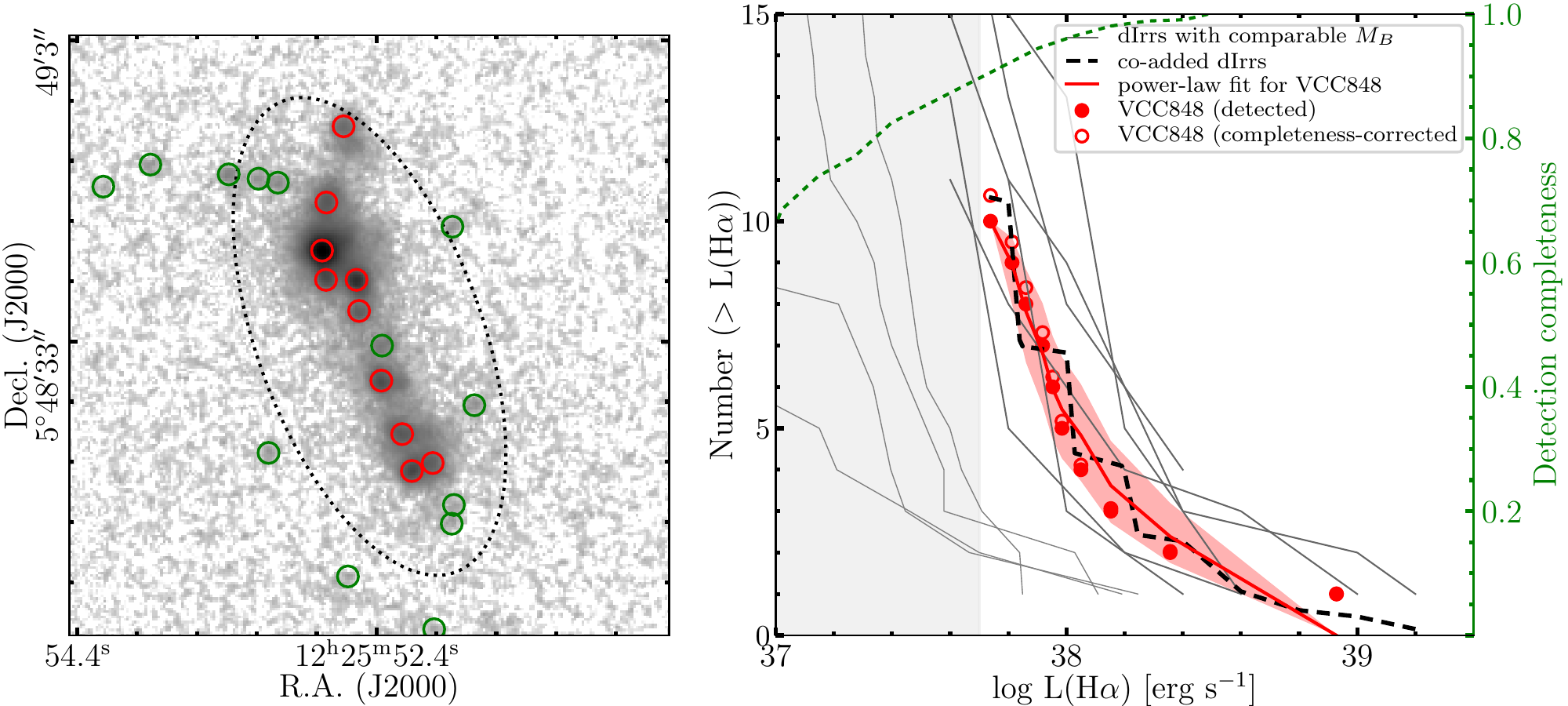}
\caption{
Central 1$\times$1 arcmin of the H$\alpha$ image of VCC 848 ({\it left}) and cumulative H$\alpha$ luminosity distribution of \HII~regions ({\it right}).\ In the {\it left} panel, 
the green and red circular apertures mark the \HII~regions brighter and fainter than the 90\% completeness limit ($>$ 10$^{37.7}$ erg s$^{-1}$), respectively.\ The big 
dotted ellipse encloses the region used for completeness estimation.\ In the {\it right} panel, the cumulative H$\alpha$ luminosity distribution of VCC 848 above the 90\% 
completeness limit is shown as red filled circles while the completeness-corrected distribution as red open circles.\ The red solid curve corresponds to a power-law fit to 
the completeness-corrected H$\alpha$ luminosity distribution of VCC 848, and the red shaded region represents the uncertainties of the power-law fit.\ The green dotted 
curve delineates the completeness limit (marked in the right y-axis) as a function of H$\alpha$ luminosities, and the gray shaded region represents the regime below the 
90\% completeness limit.\ Cumulative H$\alpha$ luminosity distributions for 10 relatively isolated dIrr galaxies whose $M_{B}$ are within $\pm$ 0.5 mag of VCC 848 
are plotted as gray curves.\ The black dashed curve corresponds to the normalized co-added distribution of the 10 dIrr galaxies.\
\label{fig_plothalf}}
\end{figure*}

To probe the impact of galaxy merging, we compare the H$\alpha$ luminosity distribution of VCC 848 to that of ordinary late-type dwarf galaxies with similar 
luminosities.\ \cite{youngblood99} and \cite{vanzee00} studied \HII~region luminosity distributions for the largest samples of relatively isolated gas-rich dIrr galaxies so far.\ 
In particular, the samples of \cite{youngblood99} and \cite{vanzee00} include 4 and 6 dIrr galaxies, respectively, that have $M_{B}$ within $\pm$ 0.5 mag of VCC 848 and 
at the same time have \HII~regions with maximum $\log L$(H$\alpha$) $\geq$ 37.7.\ We note that all of the 4 galaxies from \cite{youngblood99} happen to be fainter 
than VCC 848 while all of the 6 from \cite{vanzee00} are brighter than VCC 848.\ These 10 galaxies constitute our comparison sample.\ Cumulative H$\alpha$ luminosity 
distributions of the 10 galaxies are shown in Figure \ref{fig_plothalf}.\ In addition, we construct a composite luminosity distribution by co-adding the \HII~region catalogs 
of the 10 galaxies at $\log L$(H$\alpha$) $\geq$ 37.7, and show the resultant cumulative luminosity distribution in Figure \ref{fig_plothalf}.\ For the sake of comparison, 
the co-added distribution is normalized such that the total number of \HII~regions at $\log L$(H$\alpha$) $\geq$ 37.7 is equal to that of VCC 848.\

The cumulative H$\alpha$ luminosity distribution of VCC 848 is in reasonable agreement with the co-added distribution of the comparison sample.\ This general agreement 
is remarkable and may imply that the galaxy merging does not significantly affect the birth mass distribution of star clusters, at least at the current stage of the merger and for 
the most massive ones.\ We note that, once the full catalogs of \HII~regions down to their respective detection limit are used for the power-law fitting, the galaxies in the 
comparison sample have a median $\alpha = 1.5 \pm 0.5$, which is in fair agreement with the typical values found for dIrr galaxies and suggests that the H$\alpha$ luminosity 
distributions tend to flatten toward fainter luminosities.

Besides exploring the overall shape of the H$\alpha$ luminosity distribution above the detection limit, we can also compare the fractional luminosity contribution of the detected 
\HII~regions in VCC 848 and the comparison sample.\ In particular, at $\log L$(H$\alpha$) $\geq$ 37.7, the \HII~regions collectively contribute 47\% of the total H$\alpha$ emission 
of VCC 848, which is higher than any of the galaxies in the comparison sample ($\sim$ 15\% $\pm$ 10\%).\ The extraordinary high fractional contribution in VCC 848 is largely driven 
by the single brightest \HII~region that contributes 22\% of the total H$\alpha$ emission.\ As has been shown in Paper I, the brightest star cluster, which is associated with the 
brightest \HII~region in VCC 848, is also exceptionally luminous for the current SFR of VCC 848.\

\section{Numerical Simulations}\label{sec:simul}
We use hybrid $N$-body/hydrodynamical simulations to gain insight into the merging process of VCC 848.\ In particular, we make use of a Treecode-smoothed particle 
hydrodynamics (SPH) algorithm that largely follows the techniques described in \cite{hernquist89}, where the Treecode allows rapid calculation of gravitational accelerations 
and the SPH code allows for modeling the collisional gas component.\ The relevant Treecode-SPH parameters used in our simulations, such as the gravitational softening 
length, time-step scheme, Treecode opening angle and viscosity prescription, are the same as that adopted in \cite{smith10}.\ This Treecode-SPH code has been extensively 
used for modeling galaxy interactions and environmental effects acting on dwarf galaxies \citep[e.g.,][]{smith12a, smith12b, smith13a, smith13b, smith15}.\

We build multi-component models of a pair of otherwise isolated dwarf galaxies using the publicly available code DICE \citep{perret16}.\ In particular, each galaxy consists 
of a spherical dark matter halo with a NFW density profile \citep{navarro96}, an exponential stellar disk and, if there is gas, an exponential gas disk.\ The dark matter halo 
component consists of 400,000 particles, while the stellar and gas components consist of 50,000 particles each.\ The gas component is treated isothermally, with a constant 
sound speed of 7.5 km s$^{-1}$ (roughly equivalent to a temperature of 10$^{4}$ K).\ This has been shown to be a valid approximation of a rigorous treatment of warm 
interstellar atomic gas that has a highly efficient (inefficient) cooling efficiency at $T > $ ($\leq$) 10$^{4}$ K \citep[e.g.,][]{barnes02}.\ The gas-to-stellar disk scalelength ratio 
is fixed to 1.5, which can be compared to the typical ratio of $\sim$ 1.3 for nearby star-forming dwarf galaxies \citep[e.g.,][]{hunter12}.\ The scale height is 0.3 times the scale 
length for both the stellar and gas disks.\ Lastly, the dark matter halos have a concentration parameter $c$ $=$ 10 and their masses (M$_{\rm h}$) are linked to M$_{\star}$ 
following the relation derived from abundance matching technique \citep[e.g.,][]{guo10}.\

Given the above simulation ingredients, there is a large parameter space that involves the progenitor mass ratios, gas fractions, disk sizes, orbit geometries, etc.\ It is beyond 
the scope of the current work to conduct a complete search for all possible combinations of these parameters to match every detail of the observations.\
As elaborated below, our choice of the parameters is largely guided by the observed stellar distribution of VCC 848.\ The primary goal of our simulation is to infer the merger 
stage by qualitatively reproducing the most prominent stellar features.\ While our simulations are not meant to match the observed \HI~gas distribution of VCC 848, 
we will try to discuss the response of the gas component and its impact on the merging process that may generally apply to gas-dominated mergers of dwarf galaxies.\

\subsection{Parameter Setup}

\subsubsection{Stellar Mass and Size of Progenitor Galaxies}
An upper limit of the primary-to-secondary progenitor stellar mass ratio ($\lesssim$ 5) of VCC 848 has been estimated in Paper I.\ This upper limit was obtained by ignoring the 
contribution of the secondary progenitor to the central exponentially declining part of the stellar light distribution of VCC 848, and the mass estimate for the secondary only takes 
into account of the stellar light excess (mainly) caused by stellar shells towards large radii.\ Since the outlying stellar shells of the merger remnant consist of the least bounded 
stars stripped from the outskirts of the secondary progenitor \citep[e.g.,][]{quinn84}, it is plausible that at least a similar fraction of stars that initially inhabited smaller radii of the 
secondary has been deposited into the inner part of the remnant.\ Therefore, the true primary-to-secondary stellar mass ratio may be something close to 2 (i.e., 5-1:1+1), 
which is adopted in our simulations, with M$_{\star, {\rm primary}}$ $=$ 1.4$\times$10$^{8}$ $M_{\odot}$ and M$_{\star, {\rm secondary}}$ $=$ 0.7$\times$10$^{8}$ $M_{\odot}$.\ 
The corresponding dark matter halo mass ratio is $\simeq$ 1.3, with M$_{\rm h, primary}$ $=$ 5.3$\times$10$^{10}$ $M_{\odot}$ and 
M$_{\rm h, secondary}$ $=$ 4.2$\times$10$^{10}$ $M_{\odot}$.\ We note that the stellar to dark matter halo mass ratio decreases steeply with decreasing stellar (or halo) mass 
towards the low mass end and the dark matter halo accounts for $\geq$ 99\% of the total mass budget (and hence the gravitational force) of each galaxy in our simulations.\ This 
means that the halo mass ratio and merging timescale are not very sensitive to the exact choice of stellar mass ratio.\ For example, an increase of the stellar mass ratio by a factor 
of 2 leads to merely a $\sim$ 20\% increase in the halo mass ratio.\ Lastly, the exponential stellar disk scalelength of the simulated primary progenitor is set to 1.0 kpc while the 
scalelength of the secondary is set to be 0.8 kpc, in general accord with the luminosity-scalelength relation followed by nearby dwarf irregular galaxies \citep[e.g.,][]{hunter06}.\

\subsubsection{Orbit Geometry}
\begin{figure}[ptb]
\centering
\includegraphics[width=0.47\textwidth]{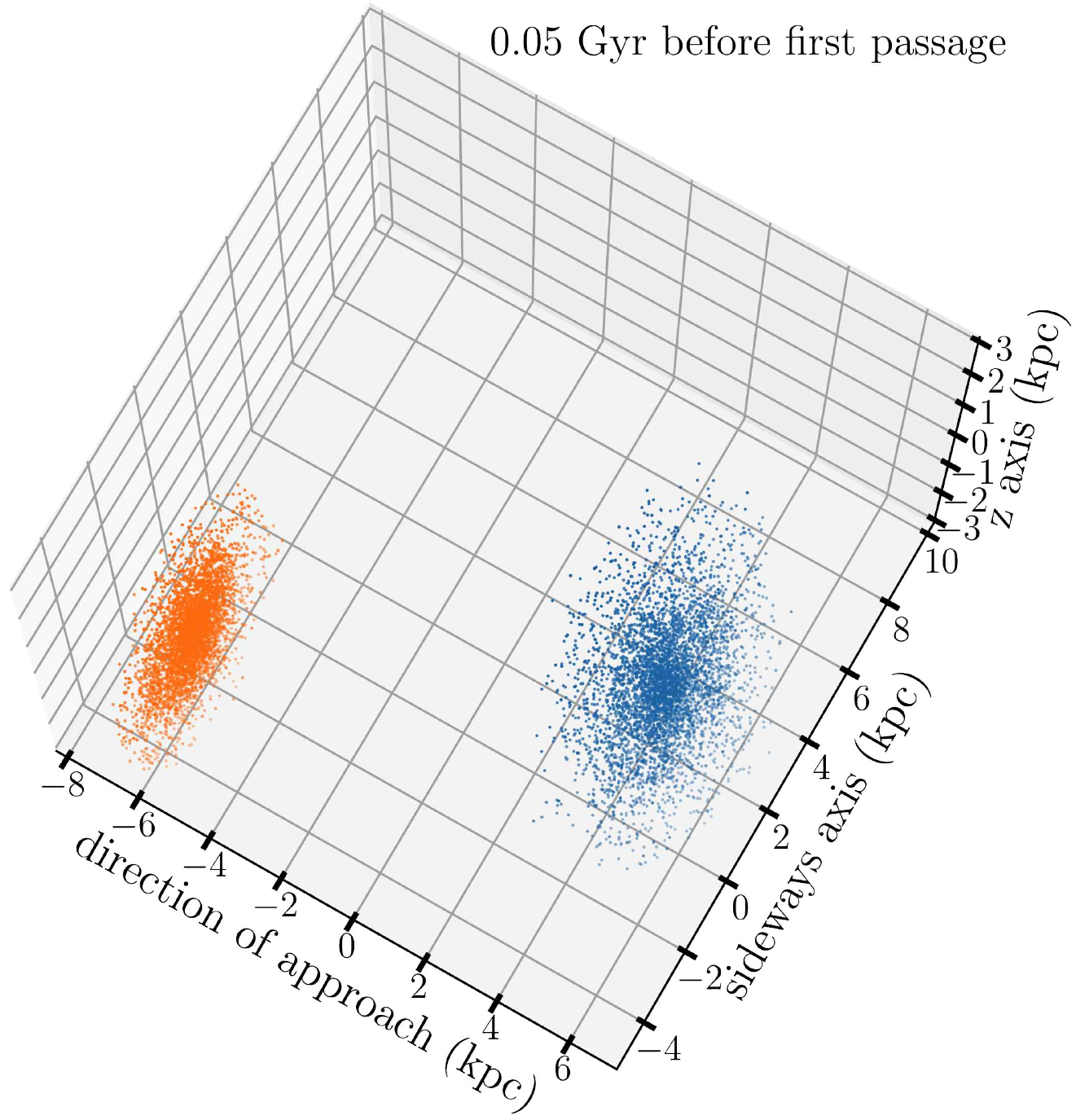}
\caption{
Illustration of the viewing direction (60$\degree$ of azimuth angle in the orbital plane, 75$\degree$ of elevation angle above the orbital plane) that leads to the best 
match between our simulation results and the stellar light distribution of VCC 848.\ The disk planes of the two progenitors (at an angle of 60$\degree$ to each other) 
are parallel with the sideways axis.\ This 3D view shows projected distributions of star particles of the primary ({\it blue}) and secondary progenitors ({\it red}) at a 
snapshot 0.05 Gyr before the first close passage.\ For clarity, only 5000 randomly selected star particles of each progenitor are plotted.
\label{fig_3dview}}
\end{figure}

The setup for the orbit geometry of the merging pair is guided by the observed stellar light distribution of VCC 848 (Paper I).\ Firstly, the remarkably extended outlying stellar 
shells that are largely aligned along the east-west direction implies a nearly radial encounter that is largely perpendicular to the line of sight direction.\ Secondly, the stellar main 
body of VCC 848 is twisted by $\sim$ 10$^{\degree}$ from the center to larger radii, which implies a moderately off-center collision.\ Thirdly, the major axis of the stellar main body 
is highly inclined with respect to the direction of shell alignment, which implies a non-coplanar collision.\ Given these observational hints, we adopt the following initial orbit  
geometry and disk orientation in our simulations.\ The two progenitor galaxies are initially 50 kpc apart, and they approach each other with a relative velocity of 50 km s$^{-1}$ 
along the collision direction and a sideways velocity of 10 km s$^{-1}$.\ Regarding the relative orientation of the galaxy disks, we ran several tests and achieve the best match 
to the observations when the disk plane of the primary (secondary) progenitor is inclined at $\simeq$ 60$^{\degree}$ (90$^{\degree}$) with respect to the orbital plane and the 
disk planes of both progenitors are parallel with the sideways direction.\ An illustration of the 3D view of the star particle distribution at a pre-encounter moment is shown in 
Figures \ref{fig_3dview}.\ 

We note that the relative approaching velocity is chosen to be smaller than the primary's escape velocity ($\sim$ 77 km s$^{-1}$ at a virial radius of $\sim$ 75 kpc).\ A larger 
approaching velocity delays the instant when the coalescence occurs and shells start to form, but makes little difference to the number of passages after coalescence nor 
to the final appearance.\ In addition, after testing with several different sideways velocities, we find that, while a sideways velocity is needed to make off-center collision and 
asymmetrical distribution of shells along the collision direction, the choice of its exact value is not crucial, due to a rapid orbital decay of the secondary caused by dynamical 
friction.\

\subsection{Results of the Simulations}
We invoke two representative simulations to discuss the progenitor properties and the merging process of VCC 848.\ Both simulations adopt a 2:1 gas-to-stellar mass ratio 
for the primary progenitor, but they differ in the gas content of the secondary progenitor.\ Specifically, the secondary progenitor is gas-free in the first simulation (hereafter S1) 
but has a 1:1 gas-to-stellar mass ratio in the second simulation (hereafter S2).\ As we will show, it is not possible to constrain the exact gas richness of the secondary, but  
we can use a 1:1 ratio in the S2 simulation to demonstrate whether the secondary is expected to be star or gas dominated.\ Figures \ref{fig_snap} and \ref{fig_snap2} show the 
projected spatial distributions of star/gas particles at selected snapshots of the S1 and S2 simulations, respectively.\ In the following subsections, we will first invoke the S1 
simulation to discuss the merging stage of VCC 848 and then the S2 simulation to discuss the impact of a gas-bearing secondary progenitor on the merger.\

\subsubsection{Reproducing the Stellar Shells of VCC 848}\label{sec:shellsimul}
\begin{figure*}[ptb]
\centering
\includegraphics[width=1\textwidth]{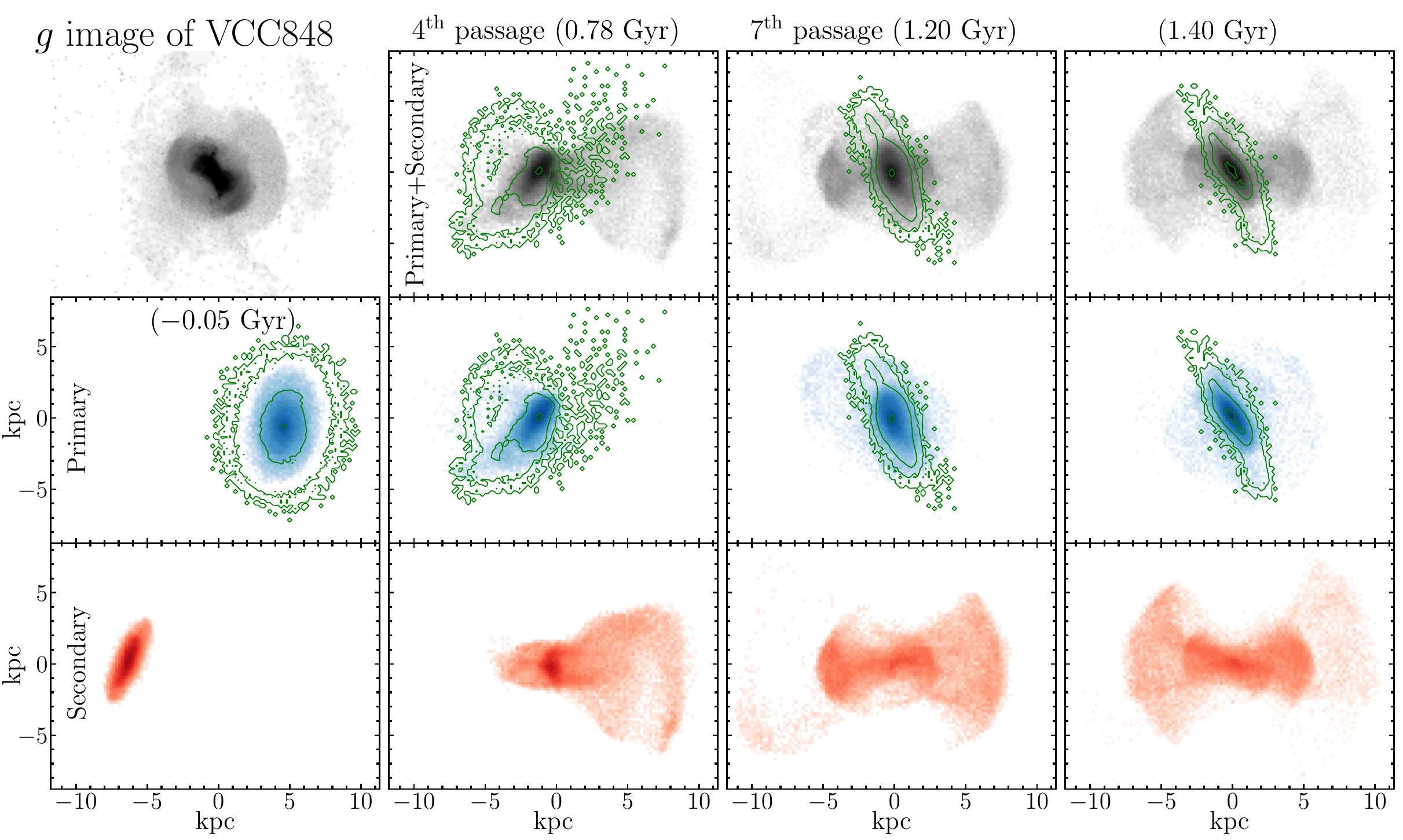}
\caption{
Projected surface number density distribution of star particles (grayscale in a logarithmic stretch) at four selected snapshots of the S1 simulation, where the primary progenitor 
has a 2:1 gas-to-stellar mass ratio while the secondary is gas-free.\ The adaptively smoothed $g$-band image of VCC 848 is shown in the top left panel for comparison 
purposes.\ For the rest of panels that show the simulation results, the time elapsed since the first close passage is indicated for each column.\ The surface number densities 
are determined at a 0.2$\times$0.2 kpc spatial resolution.\ The top panels show the primary+secondary combined stellar maps, the middle panels show the stellar maps 
of the primary, and the bottom panels show the stellar maps of the secondary.\ The gas particle surface number density distributions are overplotted in the top and middle 
panels as green contours with logarithmic intervals.\ The viewing (or projection) direction is as illustrated in Figure \ref{fig_3dview}.\
\label{fig_snap}}
\end{figure*}

The most prominent tidal features in VCC 848 are the three outlying interleaved stellar shells, among which the innermost one has significantly higher stellar surface densities, 
sharper outer edges and smaller opening angle than the other two.\ Based on the S1 simulation, the projected star particle density distributions around $\simeq$ 1.4 Gyr since the first 
close encounter of the two simulated galaxies match the above-described tidal features of VCC 848 best (Figure \ref{fig_snap}).\ We however note that the two outermost shells 
of VCC 848 have somewhat less smooth outer edges than do the ones in the S1 simulation.\ This may be ascribed to the secondary progenitor of VCC 848 being gas-bearing rather 
than gas-free as assumed in the S1 simulation (see the next two sections for more discussion).\ By the best-match moment, the progressively disassembled secondary progenitor 
has crossed through the primary for more than 7 times.\ The secondary starts to be dismantled after making the 3rd passage through the primary, and since then, one new stellar 
shell is produced every time the secondary crosses though the primary.\ While the outlying extended stellar shells are largely aligned along the east-west direction, as is observed 
in VCC 848, the direction of alignment of the shells produced near or after the best-match moment appears to be tilted toward the major-axis direction of the primary.\ This may partly 
explain the observed isophotal twist of the stellar main body of VCC 848.\

\subsubsection{Response of the Gas Component}

The gas and stellar components of the primary progenitor respond to the gravitational disturbances in a similar manner, except that the gas component develops much 
narrower tidal features than the stellar component, in general agreement with previous simulations of star-dominated galaxy mergers \citep[e.g.,][]{barnes96}.\ By the 
moment that the simulation matches the stellar light distribution of VCC 848, the gas component has largely settled into the warped central disk, with few gas particles 
associated with the extra-planar stellar tidal arms emanating from the primary, which should be attributed to orbital energy dissipation (through shock heating and radiative 
cooling) of the gas entrained in (self-)intersecting tidal arms.\ The fraction of gas particles with relatively high local (volume) number densities, as is illustrated in 
Figure \ref{fig_rhogas}, increases steadily during the course of merging.\ About 7\% of the 50,000 gas particles end up in the central 0.1 kpc in radius of the merger remnant.\ 
The enhancement of dense gas fraction may eventually results in an enhanced star formation activities in reality.\
In addition, the curve of temporal increase of dense gas fraction flattens with time, especially after the 3rd passage ($\gtrsim$ 0.6 Gyr).\ The flattening corresponds to a gradual 
weakening of the gravitational distortion of the merger remnant, due to the progressive disintegration of the secondary progenitor.\ No gas or stellar bar forms during the merging.\ 
which may be attributed to the gas dominant nature of the primary's disk (see also \citealt{athanassoula13}).\ 

There are noticeable differences between the resultant gas distribution in our simulation and the observed \HI~distribution in VCC 848.\ In particular, nearly all of the gas particles 
end up being confined to a central disk in the simulation, whereas in VCC 848, besides the high surface density \HI~gas confined to the disk plane, a significant fraction of \HI~is 
associated with the extra-planar stellar tidal arms as well (Section \ref{sec:higas}).\ We have run simulations with a initial gas disk scale length $\sim 2$ times our fiducial choice, but 
did not succeed in producing extra-planar gas arms as significant as observed in VCC 848.\ It appears that the observed \HI~distribution corresponds to an earlier merging stage (i.e., 
more violent tidal disruption) than that indicated by the stellar distributions.\ Future simulations with a comprehensive parameter-space exploration may hold the promise to achieve 
a better match to observations.\ It also remains to be seen whether starburst-driven galactic winds can significantly affect the dynamical evolution of the gas component.\ 
With that said, we note that neither of the two previous simulations of gas-rich dwarf galaxy mergers by \cite{bekki08} and \cite{starkenburg16} which include star formation and 
feedback produced prominent gas tidal tails or arms.\ In spite of the discrepancy, our discussion about the simulated gas component, including the tendency of settling towards the 
central disk, enhancement of gas densities and the impact on the stellar component (see below), should be largely valid in a qualitative sense.\

\subsubsection{Gas Content of the Secondary Progenitor}\label{sec:gasorstar}
In the S2 simulation illustrated in Figure \ref{fig_snap2}, the merging proceeds $\sim$ 80 Myr faster, and the contrast of newly formed shells relative to preceding generation of shells 
is significantly lower than in the S1 simulation.\ The faster merging in the S2 simulation reflects a faster orbital decay due to an additional mass contribution of the gas component to 
the secondary progenitor.\ During the first three passages, the secondary progenitor experiences a nearly complete gas removal and the removed gas quickly settles into the primary's 
gas disk.\ The gas removal from the secondary, which results from a dissipative collision with the primary's gas component, leaves behind a stellar system that is out of dynamical 
equilibrium and experiences a rapid expansion and dissolution.\ The contrast and decay rate of stellar shells are mainly determined by the phase space volume of the accreted 
galaxy \citep{hernquist88}, in the sense that a larger phase space volume leads to lower contrast and quicker decay.\ This explains why the shell system in the S2 simulation has 
much lower contrast and faster decay than that in the S1 simulation.\ We note that the dynamical response of the secondary's stellar system to rapid gas removal is analogous to the 
dispersal of star clusters via ``infant mortality'' \citep[e.g.,][]{hills80}.\ 

The near absence of shell contrast in the S2 simulation does not match the observations, which implies that the disk of the secondary progenitor of VCC 848 was dominated by stars 
rather than gas.\ Nevertheless, the fact that the two outermost shells of VCC 848 have less smooth outer edges than that in the S1 simulation (Section \ref{sec:shellsimul}) probably indicates that the shells in VCC 848 decay faster than in the S1 simulation and hence the secondary progenitor may not be completely devoid of (atomic) gas.\ It is however not 
possible to infer the exact gas fraction of the secondary progenitor based on the observed gas distribution, because the gas component originally belonging to the secondary, if any, 
quickly loses its identity and is phase-mixed with the primary progenitor's gas component before stellar shells start to be produced.\

In Paper I, we have shown that the average cluster formation rate of VCC 848 increased by a factor of $\sim$ 7$-$10 in the recent $\sim$ 1 Gyr, which implies that the merger 
event has triggered star formation that accounts for $\sim$ 40\% of the total stellar mass (2.1$\times$10$^{8}$ $M_{\odot}$) by assuming a constant cluster mass function.\ 
Given the current gas mass of 4.2$\times$10$^{8}$ $M_{\odot}$, the triggered star formation has consumed $\sim$ 17\% of the total pre-merger gas, which is $\gtrsim$ 3 times 
the gas inflow or high-density gas in the S1 simulation.\ In addition, an indirect constraint on the secondary progenitor's gas content may be placed by its location on the 
color-magnitude diagram.\ By adopting a ($g-i$) color of $0.6-0.7$ mag as observed for the outlying stellar shells (Paper I), which is likely an upper (redder) limit for the secondary 
progenitor as a whole, we find that the secondary is $\gtrsim 0.1$ mag bluer than the blue edge of the ($g-i$) distribution of Virgo early-type dwarf galaxies \citep{janz09,roediger17} 
for a plausible $g$ band magnitude $\lesssim$ $-15$ (Paper I).\ Above all, although being dominated by stars, the secondary progenitor of VCC 848 might contain a non-negligible 
amount of gas.\

\begin{figure}[ptb]
\centering
\includegraphics[width=0.47\textwidth]{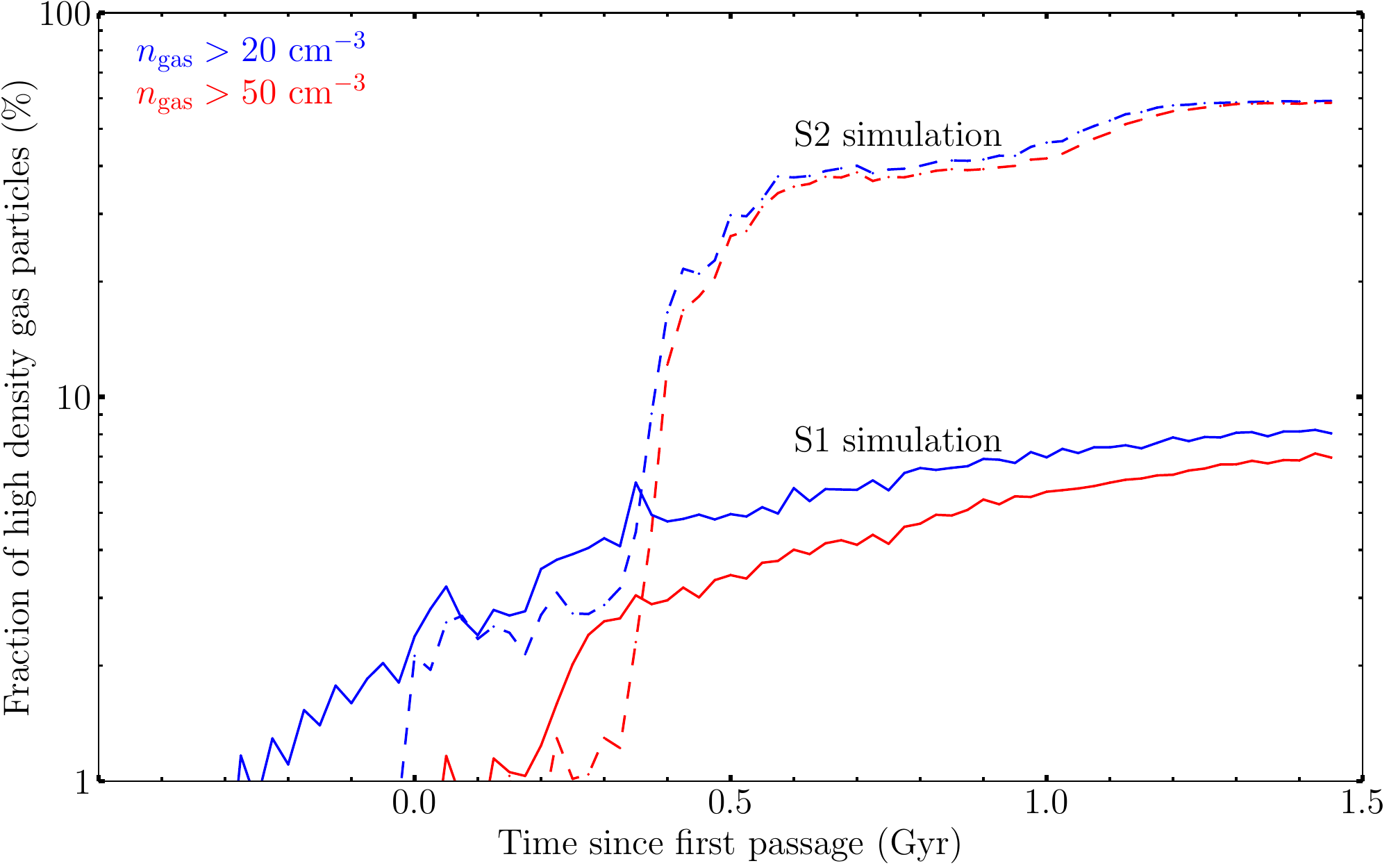}
\caption{
Temporal variation of the fraction of gas particles with relatively high local number densities $n_{\rm gas}$ in the S1 (solid curves) and S2 (dashed curves) simulations, 
respectively.\ $n_{\rm gas}$ is calculated as the average within cubic cells with 50 pc side lengths.\ Two different $n_{\rm gas}$ thresholds (in the regime of cold neutral 
medium phase) are used to illustrate the temporal variation.
\label{fig_rhogas}}
\end{figure}

\begin{figure*}[ptb]
\centering
\includegraphics[width=1\textwidth]{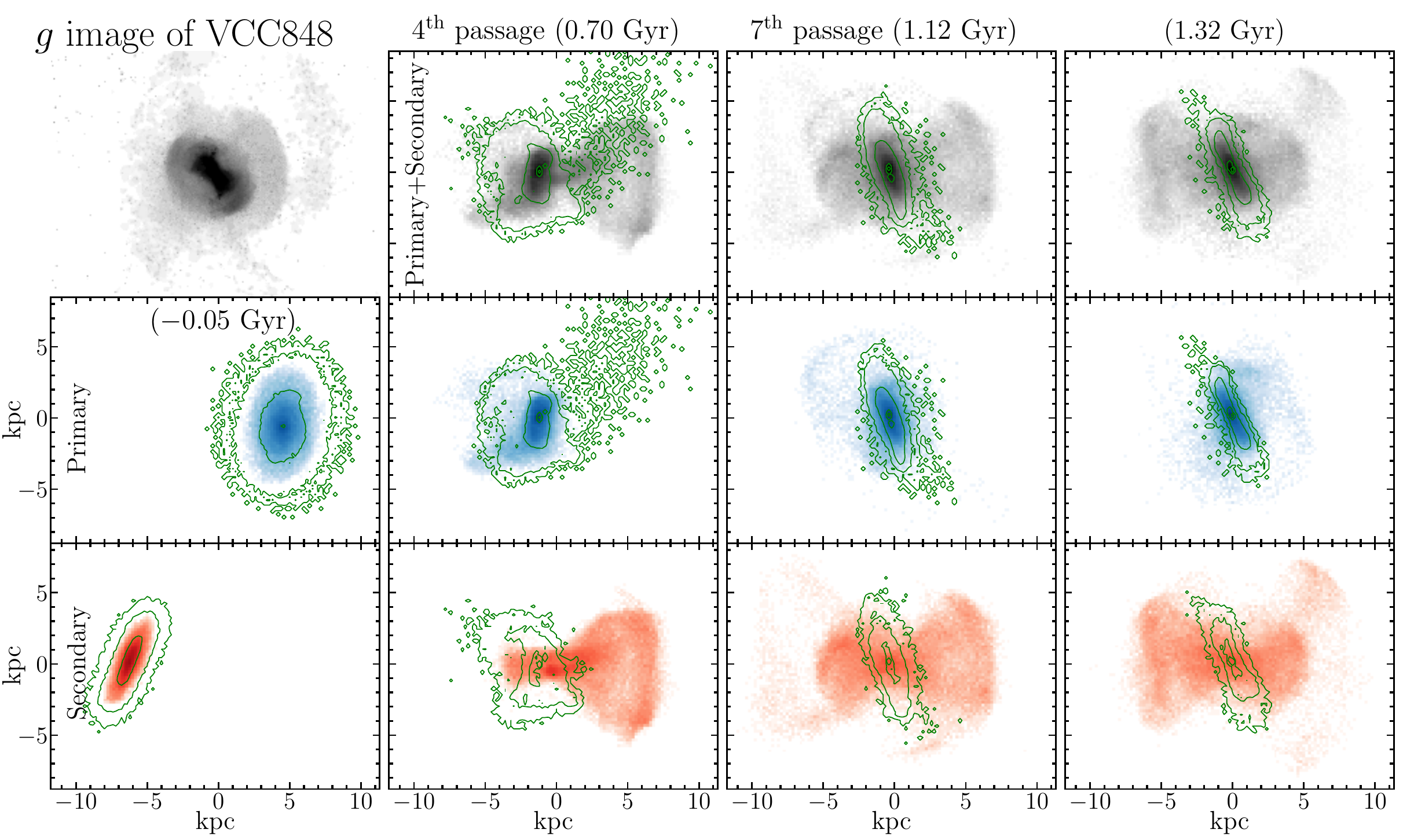}
\caption{
Same as Figure \ref{fig_snap}, but for selected snapshots of the S2 simulation, where the primary and secondary progenitors have gas-to-stellar mass ratios of 2:1 
and 1:1 respectively.\ Note that the first close passage in the S2 simulation occurs $\sim$ 80 Myr earlier than the S1 simulation.
\label{fig_snap2}}
\end{figure*}

As illustrated in Figure \ref{fig_rhogas}, gas from the secondary progenitor can dramatically affect the gas distribution of the merger remnant.\ Particularly, the S2 simulation ends 
up with an order of magnitude higher fraction of high density gas than does the S1 simulation, with $\sim$ 50\% of the gas particles flowing into the central 0.1 kpc in radius.\ 
This dramatic difference must be ascribed to the hydrodynamic interaction (e.g., ram-pressure sweeping) between the two progenitor gas disks.\ \cite{blumenthal18} pointed out 
that inward gas migration induced by the hydrodynamic interaction results in more effective coupling of the stellar and gaseous components, which favors more efficient gas inflow 
to the center driven by gravitational torques exerted by the distorted stellar disk.\ The case for VCC 848 is expected to be in between the S1 and S2 simulations.\  

\section{Summary and Discussion}

\subsection{Summary of VCC 848}
Understanding the impact of dwarf-dwarf merging on star formation and morphological evolution is of fundamental importance in the hierarchical structure formation 
paradigm.\ In Paper I, we reported the first study of VCC 848 that reveals the tell-tale signature of galaxy merger: extended stellar shells, and a significant enhancement 
of star cluster formation rate of the system during the past $\sim$ 1 Gyr.\ VCC 848 is probably by far the clearest example of star-forming galaxies formed as remnants 
of mergers between gas-bearing dwarf galaxies ($M_{\star}<$ 10$^{9}$ $M_{\odot}$).\ In the current paper, we have studied the \HI~gas distribution and star formation 
activities of VCC 848.\ We also invoke numerical simulations to probe the evolution of the stellar and gaseous components in response to gas-dominated merging between 
dwarf galaxies.\ Based on these analysis, we can draw the following summary for VCC 848.\

VCC 848 is the remnant of a merger between two gas-bearing dwarf galaxies with a primary-to-secondary stellar mass ratio of $\sim 2 - 5$ and dark matter halo mass 
ratio of $\lesssim$ 2.\ The primary progenitor is expected to be gas-dominated whereas the secondary progenitor be gas-bearing but star-dominated.\ The two progenitor 
galaxies experienced their first periapsis passage more than 1 Gyr ago, and since then a burst of star formation, as traced by the formation of super star clusters, has been 
triggered near the center of the system several hundred Myr ago (Paper I).\ SFR in the last $\sim$ $10-100$ Myr, as traced by the H$\alpha$ emission and FUV continuum, 
has dropped to a level that is consistent with being a main-sequence star-forming galaxy.\ The luminous \HII~regions in VCC 848 follow a H$\alpha$ luminosity distribution 
similar to that of ordinary star-forming dwarfs, except that the brightest \HII~region (and the associated brightest star cluster as well) in VCC 848 has extraordinarily high 
luminosity compared to its counterparts in ordinary dwarfs.\

Although VCC 848 is globally dominated by neutral atomic gas, less than 30\% of the atomic gas is associated with the stellar main body which contains nearly all of the 
total SFR.\ Molecular gas mass inferred from the SFR seems to dominate over the atomic gas mass in the central $\sim$ 1.5 kpc.\ It is conceivable that a large amount of 
atomic gas has been transformed to the molecular phase that fuels the active star formation activities in the recent past.\ By contrast, a significant fraction of 
the atomic gas entrained in tidal arms reaches column densities as high ($\gtrsim$ 10$^{21}$ cm$^{-2}$ or 10 $M_{\odot}$ pc$^{-2}$) as that in the main body but with 
negligible associated star formation.\

The observed stellar distribution of VCC 848 implies a nearly radial and non-coplanar encounter between two dwarf galaxies whose disks are highly inclined with respect to 
the orbital plane.\ Early simulations of spiral-spiral mergers \citep[e.g.,][]{mihos96} suggest that non-coplanar encounters do not result in gas inflow and starburst activities 
as strong as coplanar encounters do.\ More recent simulations by \cite{blumenthal18} find an intricate dependence of gas inflow on both encounter parameters and progenitor 
properties.\ In particular, they find that a higher gas content or larger gas-to-stellar disk size ratio generally leads to less effective gas inflow.\ One may therefore be tempted to 
deduce that dwarf-dwarf mergers are not as effective as spiral-spiral mergers in driving gas inflow and centralized starburst in general.\ Nevertheless, our simulations tailored to 
VCC 848 suggests that, while the merger of a gas-dominated primary with a gas-free secondary results in weak gas inflow ($\sim$ 5\%), the merger with a gas-bearing secondary 
can drive a substantial gas inflow (up to $\sim$ 50\%) to the center of the remnant, even in the seemingly disfavored encounter geometry.\ The substantial gas inflow seen in our 
simulations with a gas-bearing secondary is in agreement with SPH-based simulations by \cite{bekki08} and \cite{watts16}.

Our dynamical analysis of the \HI~velocity field suggests that the merger event has not built up a central region as compact as typical compact dwarfs with centralized 
starburst (Section \ref{sec:rotcurvslope}).\ This may be explained either by a relatively low gas content in the secondary progenitor (thus ineffective gas inflow) according 
to our simulations, or by a substantial gas outflow driven by stellar feedback following centralized starburst in early stages of the merger.\ Stellar feedback is expected to 
be more effective in driving galactic outflows in dwarf galaxies than in massive galaxies.\ \cite{kim09} ran the first adaptive mesh refinement (AMR) simulation of a 
gas-rich dwarf-dwarf merger by incorporating a high-resolution treatment of the multiphase interstellar medium and shock-induced star formation, and they found that 
more than 90\% of the gas is expelled by stellar feedback.\ In VCC 848, the centralized starburst triggered at early stages of the merger may have driven substantial 
galactic outflows that evacuate the gas that once flowed into the central region.\

\subsection{Reflection on the outcome of dwarf-dwarf mergers}

A general conclusion from early simulations \citep[e.g.,][]{mihos96} is that mergers of galaxies lacking compact bulges (as is generally true for normal star-forming dwarf 
galaxies) develop substantial gas inflows and induce intense starburst activities at much earlier merging stages than those with bulges.\ This is because bulges act to 
suppress the development of strong non-axisymmetric structures (e.g., bars) that helps driving gas inflows until the final coalescence of mergers.\ Our result that the 
most intense starburst activities in VCC 848 happened near the center at earlier stages of the merger appear to be in line with these simulations.\ It is noteworthy that 
a recent study by \cite{privon17} found widespread starburst activities in a well-separated pair of dwarfs that appear to be at an early stage of merging.\ In addition, the 
lack of a compact core in VCC 848 is also in line with the general expectation that stellar feedback may drive substantial galactic outflows that act to suppress the formation 
of stellar bulges in low-mass galaxy mergers \citep[e.g.,][]{ubler14}.\ 

If these findings are verified to be generally true for dwarf-dwarf mergers, it would negate the possibility that mergers between (normal) low surface brightness star-forming 
dwarfs can result in compact dwarf remnants.\ The observed compact dwarfs with centralized starburst activities, if being involved in mergers, may be formed from 
progenitors that already had bulge-like compact cores prior to the merger event.\ Mechanisms such as gas accretion from the cosmic web \citep[e.g.,][]{sanchez14} 
and disk instabilities \citep[e.g.,][]{elmegreen12} may have contributed to formation of pre-existing compact cores.\ Pre-existing compact cores may help sustain  
strong starburst activities when their host galaxies are involved in galaxy mergers.\

It has long been recognized that gas plays a disproportionately large role in interacting and merging galaxies.\ The usually gas-dominated nature of star-forming dwarf galaxies 
means that dwarf-dwarf mergers may not be simply the scale-down counterparts of (star-dominated) spiral-spiral mergers in the local universe.\ Although star-forming dwarf 
galaxies in the local universe can reach similarly high gas richness as the observed high-redshift massive star-forming galaxies, gas in the disks of local dwarfs is primarily in 
a warm diffuse atomic phase whereas it is mainly in a cold clumpy molecular phase for high-redshift massive galaxies \citep[e.g.,][]{tacconi10}.\ As has been pointed out by 
\cite{fensch17}, local gas-rich dwarf mergers may induce more significant enhancement of gas inflow and star formation than mergers of high-redshift massive clumpy galaxies 
{\it in a relative sense}.\ A complete understanding of the outcome of dwarf-dwarf mergers requires detailed analysis of representative samples of dwarf systems at both early  
\citep[e.g.,][]{stierwalt15} and late stages of merging.\ On the simulation side, a more realistic treatment of the multiphase interstellar medium, star formation and feedback process, 
will be indispensable in unraveling the physical processes governing the evolution of dwarf-dwarf mergers in general.

\begin{acknowledgements}

We thank the anonymous referee for her/his constructive comments that improved the manuscript.\
We acknowledge support from the National Key R\&D Program of China (2017YFA0402702), the NSFC grant (Nos.\ 11973039 and 11421303), and the 
CAS Pioneer Hundred Talents Program.\ SP acknowledges support from the New Researcher Program (Shinjin grant No.\ 2019R1C1C1009600) through 
the National Research Foundation of Korea.\ SHO acknowledges a support from the National Research Foundation of Korea (NRF) grant funded by the 
Korea government (Ministry of Science and ICT: MSIT) (No. NRF-2020R1A2C1008706).\ YG's research is supported by National Key Basic Research and 
Development Program of China (grant No. 2017YFA0402700), National Natural Science Foundation of China (grant Nos.\ 11861131007, 11420101002), and 
Chinese Academy of Sciences Key Research Program of Frontier Sciences (grant No.\ QYZDJSSW-SLH008). 

\end{acknowledgements}

\software{{\sc casa} \citep{mcmullin07}, {\sc 2dbat} \citep{oh18}, {\sc SExtractor} \citep{bertin96}, DICE \citep{perret16}}

\restartappendixnumbering

\appendix

\section{\HI~Channel maps of VCC 848}
\begin{figure*}[pt]
\centering
\includegraphics[width=0.93\textwidth]{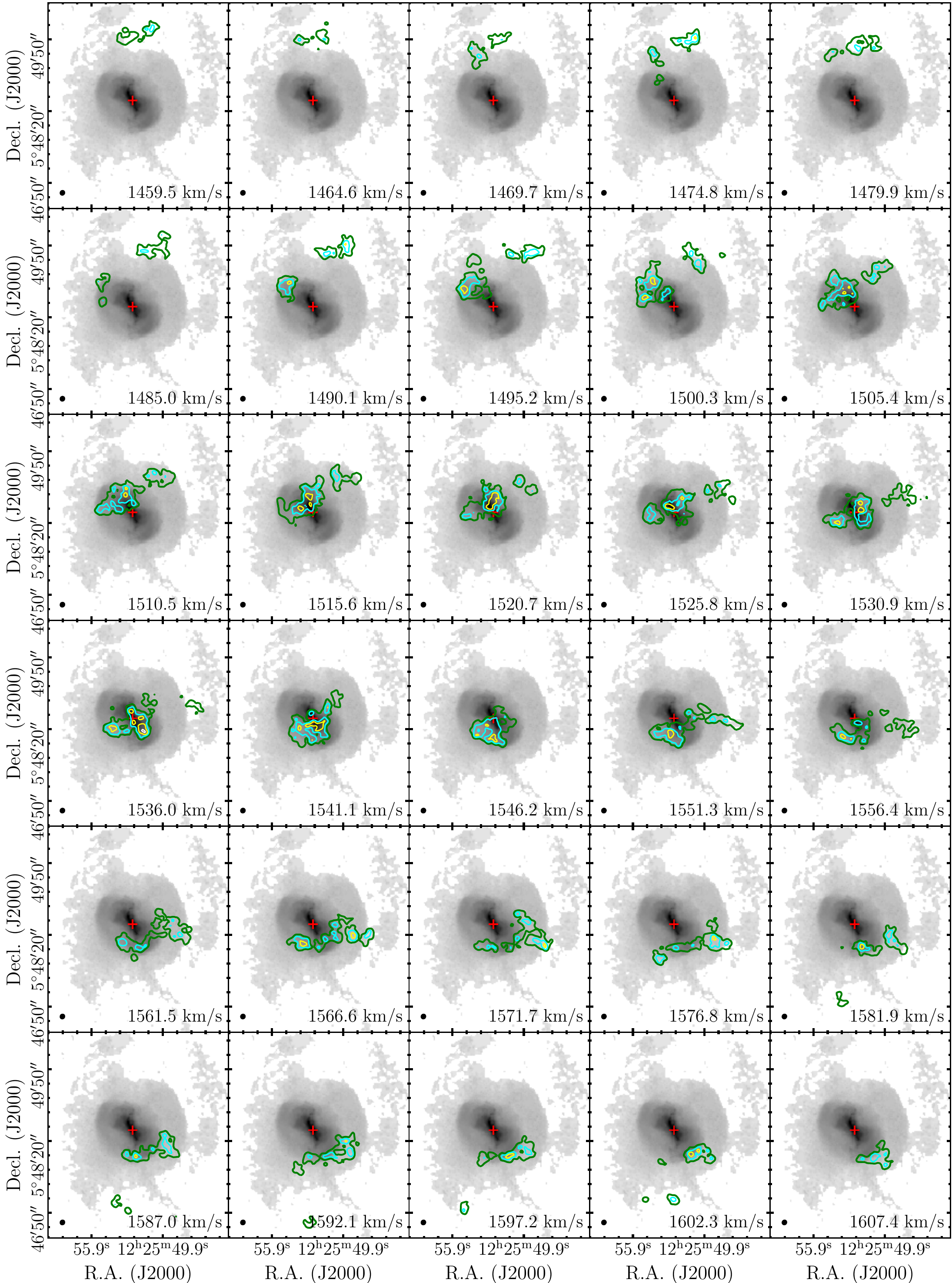}
\caption{
\HI~channel maps of VCC 848 contoured on the $g$-band image.\ The original \HI~cube is spectrally re-binned to a 5.1 km s$^{-1}$ channel width, and the $g$-band 
image is re-binned to a pixel size of 1\farcs5 in order to match the \HI~data cube.\ The contours are drawn at levels of 3 ({\it green}), 6 ({\it cyan}), 9 ({\it yellow}) and 
12 ({\it white}) times the single-channel noise of 0.64 mJy beam$^{-1}$.\ The beam size (7\farcs1 $\times$ 6\farcs6, PA = 56.78$^{\circ}$) is indicated in the bottom left 
corner of each panel.\ The photometric center is marked by a red plus symbol in each panel.
\label{fig_hichan}}
\end{figure*}

The \HI~channel maps of VCC 848 are contoured on the $g$ band image in Figure \ref{fig_hichan}.\

\end{document}